\documentclass[12pt]{article}

\usepackage{epsfig}
\usepackage{cite}
\usepackage{inputenc}
\inputencoding{latin1}

\newcommand{\EPJC}[3]{{\it Eur.~Phys.~J.} {\bf C#1} ({#3}) {#2}}

\newcommand{\JMP}[3]{{\it J.~Math.~Phys.} {\bf #1} ({#3}) {#2}}
\newcommand{\JPG}[3]{{\it J.~Phys.} {\bf G#1} ({#3}) {#2}}
\newcommand{\ZETF}[3]{{\it Zh.~Eksp.~Teor.~Fiz.} {\bf #1} ({#3}) {#2}}

\newcommand{\PLB}[3]{{\it Phys.~Lett.} {\bf B#1} ({#3}) {#2}}
\newcommand{\YF}[3]{{\it Yad.\ Fiz.} {\bf #1} ({#3}) {#2}}
\newcommand{\SJNP}[3]{{\it Sov.~J.~Nucl.~Phys.} {\bf #1} ({#3}) {#2}}
\newcommand{\JETP}[3]{{\it Sov.\ Phys.\ JETP} {\bf #1} ({#3}) {#2}}
\newcommand{\JETPL}[3]{{\it JETP~Lett.} {\bf #1} ({#3}) {#2}}
\newcommand{\NPB}[3]{{\it Nucl.~Phys.} {\bf B#1} ({#3}) {#2}}
\newcommand{\PRD}[3]{{\it Phys.~Rev.} {\bf D#1} ({#3}) {#2}}
\newcommand{\PRL}[3]{{\it Phys.~Rev.\ Lett.} {\bf #1} ({#3}) {#2}}
\newcommand{\ZPC}[3]{{\it Z.~Phys.} {\bf  C#1} ({#3}) {#2}}
\newcommand{\PR}[3]{{\it Phys.~Rep.} {\bf #1} ({#3}) {#2}}

\newcommand{\HEPPH}[1]{{\it hep-ph/#1}}
\newcommand{\HEPEX}[1]{{\it hep-ex/#1}}

\begin{document}

\begin{titlepage}

  \begin{flushright}
    HEP-PH/0004022\\
    DESY 00-054\\
    April 2000\\
    Revised: July 2000
  \end{flushright}
  \begin{center}

    \vskip 6mm
    {\LARGE\bf A perturbative treatment of double gluon exchange in 
      $\gamma^*$-proton DIS}
    \vskip 8mm

    {\large Hamid Kharraziha}\footnote{Supported by the EU Fourth Framework 
      Programme "Training and Mobility of Researchers", Network 
      "Quantum Chromodynamics and the Deep Structure of Elementary Particles",
      contract FMRX-CT98-0194 (DG 12 - MIHT).}\\[0.25cm]
    {\it II. Institut f\"ur Theoretische Physik}\\
    {\it Universit\"at Hamburg}\\
    {\it D-22761 Hamburg, Germany}\\
    hamid.kharraziha@desy.de\\
    Tel: +49 40 8998 2234\ \ \ \ Fax: +49 40 8998 2267 

  \end{center}
  \vskip 6mm
  \begin{abstract}
    \noindent
    A new model for the exchange of two gluons between the virtual photon
    and the proton, in non-diffractive deeply inelastic 
    electron-proton scattering, 
    is developed and studied. This model relies on a perturbative calculation,
    previously applied to diffraction, and a general result from Regge
    theory. As a first application of the model, we study corrections to the 
    momentum transfer to the quark-anti-quark pair, at the photon-vertex. We 
    find a significant enhancement of the cross-section at $\sim Q^2$ momentum 
    transfers, and large negative corrections for small momentum transfers. 
    The implication of this result for jet-distributions measured at HERA, is
    discussed.  
  \end{abstract}
\end{titlepage}
\section{Introduction}
In an additive parton model description, the rapid growth of the 
electron-proton cross-section with the total energy $s$, is
interpreted as a growth in the number of partons (mainly gluons), 
inside the proton. When the density of the partons is sufficiently
large, it is generally expected that they will start to self
interact (screening corrections), and that the quark probe, which 
measures the number of gluons, will interact with several gluons 
simultaneously (shadowing). Such corrections are expected to slow down
the energy growth of the cross-section, in a way that the Froissart 
bound, $\sigma\propto {\rm log}^2(s)$ \cite{frois}, is not violated.

It has, however, been shown, for large enough virtualities of the photon,
$Q^2>1~{\rm GeV}^2$, where perturbation theory is applicable, that the
structure function measured at HERA, is described both by models that do not
include multiple interaction effects \cite{DGL1}, and by models where 
multiple interactions have a significant contribution (See e.g. \cite{DGL2}). 
The main theoretical uncertainty, is the non-perturbative nature of the
proton. Different perturbative behaviour can describe the same structure 
function, by making different non-perturbative assumptions for the proton.

One way of settling this ambiguity, is to study more exclusive observables,
e.g. properties in the final-state distribution, in addition to the structure
function. A striking example of this is the observation of rapidity gap
events. The appearance of large regions in rapidity with no hadrons in the
final state, are generally considered to be a signal of the exchange of a
colour singlet state between the proton and the probe. In a perturbative
description, this is only possible by the exchange of, at least, two gluons. 
Thus, the rapidity gap events, indicate the significance of multiple
interactions. 

It is nowadays, well established that, in many cases, multiple interactions have
a significant effect also on the non-diffractive final states. This became
first evident, in proton-anti-proton scattering. It was observed that multiple
interactions considerably increase the fluctuations in the final-state 
multiplicities of charged hadrons. Assuming universality for the hadronisation
process, it was shown \cite{torbjorn} that multiple interactions are 
necessary for the  description of high multiplicity distributions, measured
by the UA5 collaboration for proton-anti-proton scattering at 540 GeV
\cite{UA5}. Direct measurements of the cross-section for double
independent parton scatterings, have been made, from the appearance of two 
uncorrelated pairs of balanced high-$p_\perp$ jets \cite{ppmulti,CDF} 
(In \cite{CDF}, the measurement was made on 3 jets + photon final-states.). 

The same phenomenon is observed in the case of electron-proton scattering at
very low virtualities of the photon. The hadronic component of the quasi real 
photon, enables a description of multiple interactions, similar to the case of
hadron-hadron scatterings. Again, one picks out two (or more) partons from
each projectile, and describe the multiple interaction process as
simultaneous $2\rightarrow 2$ scatterings \cite{phmulti}. In this way, the 
inclusive distribution of each scattering becomes the same as for single
scattering, or in other words, inclusive final-state distributions are not
changed. In photon-proton scattering, a clear signal for multiple interactions
were found in the form of the so called "Pedestal effect" with an enhanced 
low-$p_\perp$ activity in the surrounding of high-$p_\perp$ jets \cite{pedes}.

In this article, the {\it independent} multiple scatterings,
described above, is complemented with a model which relies on a perturbative
treatment of the coupling of the photon, to two gluons inside the proton. 
This model results in a different kind of final-state, compared
to the case of two independent scatterings. It does not necessarily lead to an
increase in the jet activity in the photon direction, but instead, it leads to
a change of the inclusive jet distributions. It is perhaps also more relevant
for virtual photons, which should not be expected to have non-perturbative
hadronic properties. 

In Section \ref{model}, the basic ideas behind the model are described. One of
the main ingredients is a perturbative treatment of the photon vertex which is
relevant for, and has previously been applied to, diffractive events in 
electron-proton scattering \cite{Diff}. A general result from Regge theory 
(The AGK cutting rules \cite{AGK}), is then used
to relate this perturbative calculation, to the corresponding double gluon
exchange corrections to non-diffractive DIS events. 
By following this strategy, the relevant formulas (Eqs. (\ref{s2}) and
(\ref{s1})) are derived in Section
\ref{logs}. In Section \ref{prota}, the four-gluon impact factor of the
proton is determined, by phenomenological arguments. 

In Section \ref{num}, the model is used to make numerical estimates for 
the effects on the momentum transfer
($p_\perp$) between the proton and the quark-anti-quark pair at the photon
vertex. We find that the differential cross-section $d\sigma/dp_\perp^2$, for 
$p_\perp^2\sim Q^2$ is increased with about 10\%, for $Q^2=10$~${\rm GeV}^2$ and
$x_B=0.0001$, relevant for the HERA region, and that this correction grows
rapidly towards lower values of $x_B$ and $Q^2$. It is however pointed out that
the uncertainties in this estimate are large, in this first treatment.
In the low-$p_\perp$ region, we find large negative corrections. This is
interpreted as a signal for saturation in the $p_\perp$-distribution.

In the Appendix, we make a consistency check for the validity of the
AGK-rules, in the sense they have been used for the model. This is however
not a full QCD calculation.

It should be mentioned, that the same kind of physics that is discussed here, 
can be treated inside a formalism that has been developed for radiation 
process in medium \cite{LPM1,LPM2}. So far, this study has been concentrated
on the effect of energy loss in medium, but recently \cite{LPM2}, it was
extended to also cover effects on transverse momenta distributions. Explicit 
studies in this respect for e-p scattering, has however not yet been done.

\begin{figure}
\psfig{figure=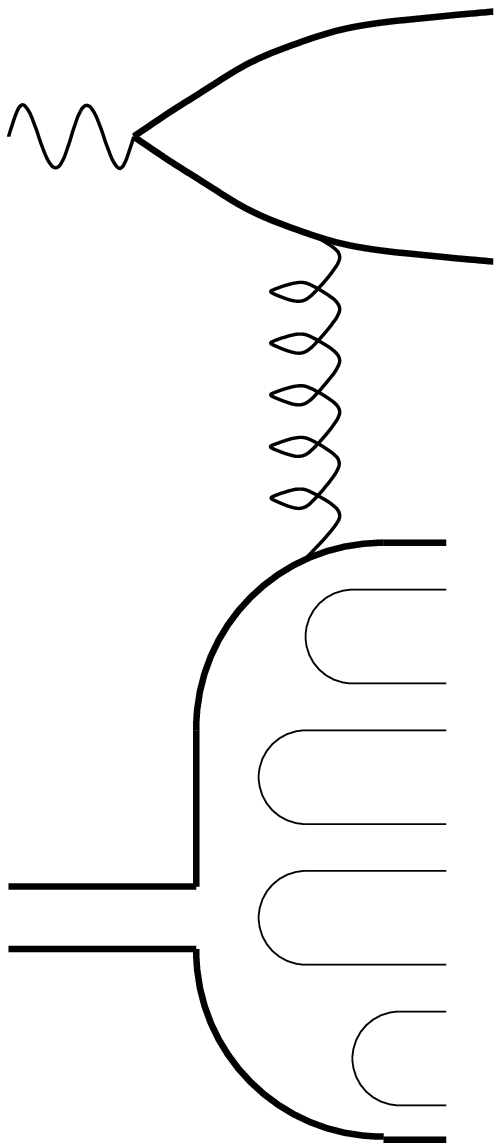,width=3cm}
\hskip 2cm
\psfig{figure=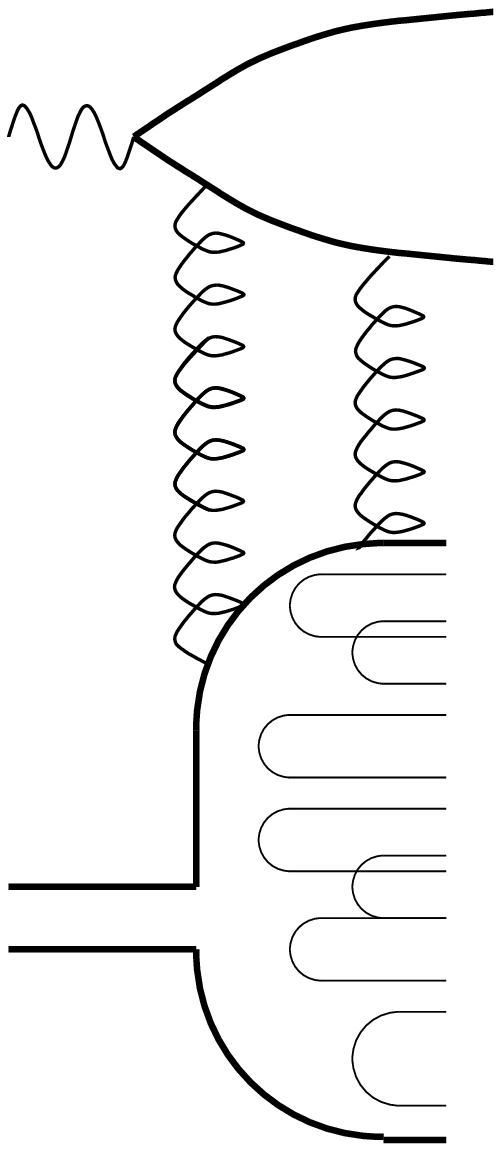,width=3cm}
\hskip 2cm
\psfig{figure=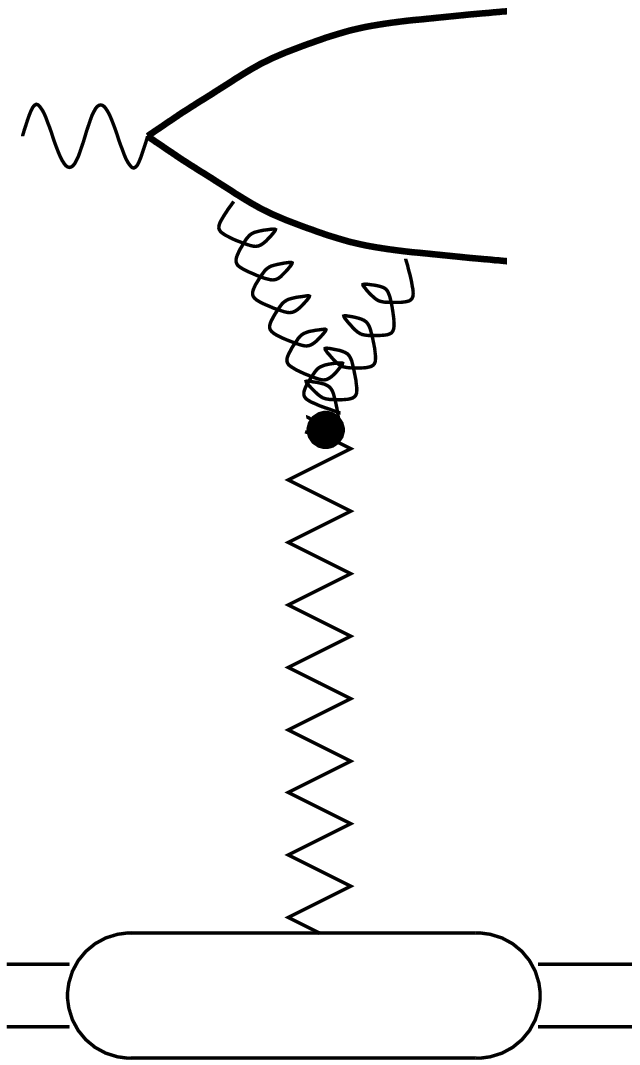,width=4cm}
\\
\begin{center}
(a)\hskip 5cm
(b)\hskip 5cm
(c)
\end{center}
\caption{(a) Normal DIS: A single gluon is exchanged between the quark-pair
  (above) and the proton (below). The proton breaks up into a high-mass 
  hadronic state. (b) Non-diffractive double gluon exchange: The hadronic
  state becomes more dense. (c) Diffractive event: A pomeron is exchanged. It
  couples to the quark-pair as a system of two gluons in colour singlet.}
\label{procfig}
\end{figure}

\begin{figure}
\psfig{figure=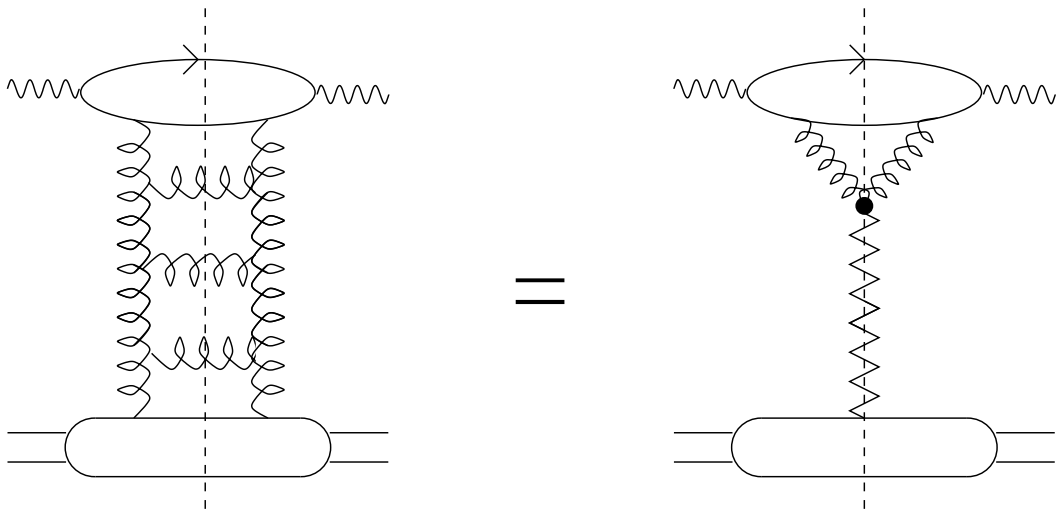,width=14cm}
\caption{The BFKL ladder and its representation as one-pomeron exchange.}
\label{ladderfig}
\end{figure}

\begin{figure}
\begin{center}
\psfig{figure=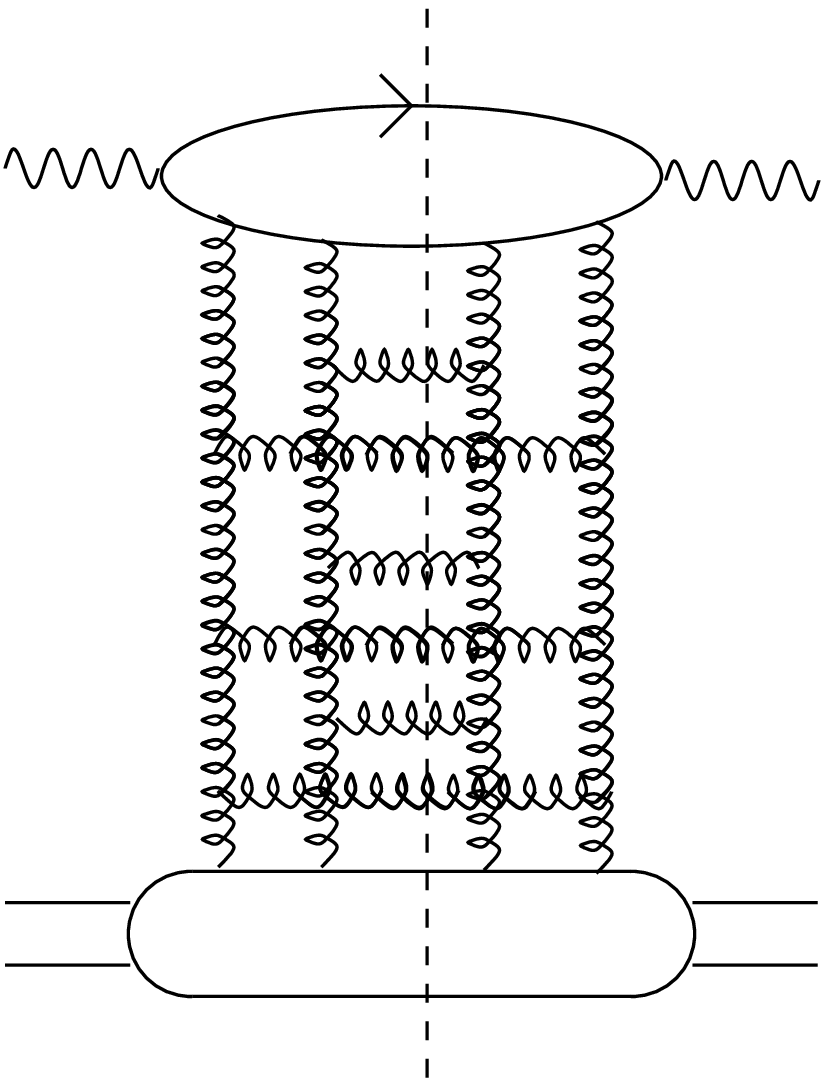,width=5cm}
\hskip 3cm
\psfig{figure=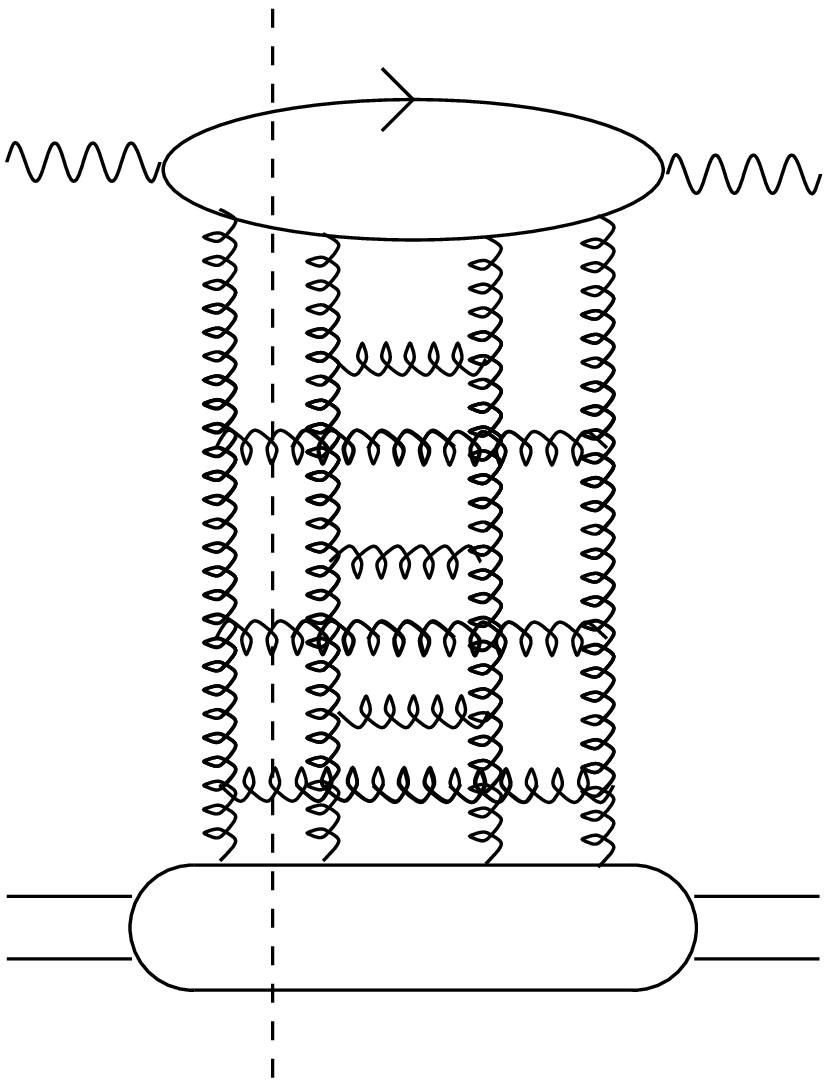,width=5cm}
\\
(a)\hskip 9cm
(b)
\end{center}
\caption{Examples of two-ladder exchange, with different final states: (a)
Both ladders are cut and the final-state distribution is doubly dense,
compared to normal DIS.
(b) One ladder is cut. This gives the same final-state distribution as the
leading contribution.}
\label{ladder2fig}
\end{figure}

\section{A model for perturbative double gluon exchange}
\label{model}
The details of the present model are outlined in the following two
sections. Here, we discuss the physical problem in general, and 
describe the main assumption that are used in the model.

In the region of low $x_B$ and moderate $Q^2$, the leading ${\rm log}(s)$
(LLs) calculations of BFKL \cite{BFKL} are expected to be
valid, and gave in fact, a qualitatively correct prediction for the growth of
the cross-section, towards lower $x_B$-values, $\sigma\propto
x_B^{-\lambda}$.
(It should be mentioned that also models based on 
DGLAP evolution give a good description of the strong rise of the 
structure function \cite{DGL1}.)
In this frame work, the leading contribution to the cross-section is given by 
the exchange of a colour singlet gluonic ladder between the proton and a quark
loop at the photon vertex (See Fig. \ref{ladderfig}). The rungs in the
ladder appear as gluon emissions into the final-state. After the 
hadronisation process, one will then
end up with a hadronic final-state with a very large mass, $W^2=Q^2/x_B$.
The BFKL prediction for the value of the exponent 
$\lambda=4\bar{\alpha}{\rm log}(2)\approx 0.55$, for 
$\bar{\alpha}\equiv N\alpha_s/\pi=0.2$ with $N=3$ colours, is observed
to be too large compared to measurements of the structure function. 
Sub-leading corrections are expected to lower this value to around 
$\lambda=0.3-0.4$ (See e.g. \cite{LDC} and \cite{salam}). For this
cross-section, down to the next-to-leading order, we will use the
notation $\sigma_1^{(2)}$, since it corresponds to the
process that the quark pair absorbs a single gluon from the proton and
has its contributions given from diagrams with two reggeised 
gluons coupled to the quark loop.

The multiple interaction effects that we are interested in here, 
appear first at the sub-sub-leading level and are currently not fully
calculated. For double gluon exchange\footnote{Note that we are 
here discussing
the  underlying physical process of two exchanged gluons, however, 
  the processes of double gluon exchange and the
  correction to single gluon exchange, have their contribution given from
  Feynman diagrams with four exchanged gluons. We use the notion 'single' and
  'double' when discussing the physical process, and numbers e.g. 'two-', 
  'four-' for the discussion of Feynman diagrams.}
(With the cross-section given from diagrams with four reggeised 
gluons coupled to the quark loop.) 
  one can however expect contributions with two
exchanged ladders, which would lead to a faster energy growth
than the leading contribution. We will use results from Regge theory, 
as a way to determine just these
fast growing contributions at the sub-sub-leading level.

In the language of Regge theory, the gluon ladder which appears in the
leading BFKL contribution corresponds to the exchange 
of a pomeron, with the final-state cut going across it.
In Regge theory, one can also discuss the possibility of the exchange
of several pomerons. Two-pomeron
exchange, with the final-state cut going in between the two pomerons,
has been successfully used for the description of diffractive
events in electron-proton scattering \cite{Diff}. There, the
pomeron and its coupling to the proton, was determined by phenomenological
arguments, while, at the photon vertex, the pomeron was assumed to couple
 as two gluons in the colour singlet state.
The leading part of the amplitude for pomeron exchange is expected to be purely
imaginary. According to the Cutkosky rule \cite{cut}, this means that it is, 
to leading order, given by its s-discontinuity, 
$$
{\cal A}_{\rm diff}\approx
{\rm iIm} {\cal A}_{\rm diff}=\frac{\rm i}{2}{\rm disc}_s {\cal A}_{\rm diff}
$$
The diffractive cross-section, which we will denote $\sigma_0$, is in 
this way determined from a triple discontinuity and the calculations 
are considerably simplified.

In Regge theory, the two-pomeron exchange gives two other types of
contributions, but with non-diffractive final-states. These are given
by the possibilities that the cut goes through one or both of the
pomerons. We will base the modelling of double gluon exchange and a 
higher order correction to single gluon exchange on these
non-diffractive contributions of two-pomeron exchange. The
contributions to double gluon exchange, the process that the quark 
pair absorbs two gluons, are given from diagrams like the one in
Fig.~\ref{ladder2fig}a with two cut pomerons. The contribution from
diagrams with one cut pomeron (Fig.~\ref{ladder2fig}b), correspond
to the higher order correction to single gluon exchange. 

Throughout in this article, we denote the cross-section of single gluon
exchange $\sigma_1$ and the cross-section of double gluon exchange 
with $\sigma_2$. Single gluon exchange has two separate
contributions, $\sigma_1=\sigma_1^{(2)}+\sigma_1^{(4)}$, where
$\sigma_1^{(2)}$ is the BFKL cross-section with two reggeised 
gluons coupled to the quark loop, while $\sigma_1^{(4)}$ is the 
sub-sub-leading correction, with four reggeised gluons. (Some sub-leading
corrections in the gluon ladder are also included by using the lower
value of the exponent $\lambda$.) The diffractive
cross-section is denoted $\sigma_0$. For the sake of discussion,
a tilde will be used for the cross-sections defined in Regge
theory. Thus, we will denote the two-pomeron exchange cross-section
with $i$ cut pomerons with $\tilde{\sigma}_i$. However, in the present
approach, each of these will be identified with the respective untilded
cross-section (With the same sub-script). 

For the non-diffractive contributions of two-pomeron exchange,
i.e. $\sigma_2$ and $\sigma_1^{(4)}$, it is not certain that the use 
of the Cutkosky rule gives the complete answer, since the amplitudes 
(The diagram on either side of the cut in Fig. \ref{ladder2fig} a and b.) 
can have real contributions, also at the level, in the LLs expansion, 
that we are interested in. We will resolve this uncertainty, by using a
result in Regge theory. In an article by Abramovski, Gribov and 
Kancheli \cite{AGK}, it was shown that the relative size of
the contributions of two-pomeron exchange, but with different number
of cut pomerons is given by
\begin{equation}
\label{agk}
\tilde{\sigma}_0:\tilde{\sigma}_1:\tilde{\sigma}_2=1:-4:2.
\end{equation}  
The main
ingredient in their proof was factorisation into one, approximately 
imaginary, signature factor for each pomeron, and a single vertex 
function for the coupling of the
two pomerons to the external particles. However, they used an
underlying $\phi^3$-theory of scalar particles, in order to show that
the vertex functions are not changed in the different cuttings. We will
make the non-trivial assumption that the AGK rules are valid also in 
our case. This
allows us to use the diffractive calculation, where the Cutkosky rule is
applicable, to determine the other two contributions $\sigma^{(4)}_1$ 
and $\sigma_2$.

As described above, the diffractive cross-section is given from
diagrams with one pomeron on each side of the final-state cut. Each
pomerons couples to the quark loop as two gluons in the colour singlet
state and with the transverse momenta relatively well balanced. For 
the case of double gluon exchange $\sigma_2$ (Fig. \ref{ladder2fig}a)
one also has two gluons that couples to the quarks, on each side of the 
final-state cut, but they arise from two different pomerons. 
They are not necessarily in the colour singlet state, and the
momenta can be rather unbalanced, due to perturbative QCD
radiation. Finally, the higher order correction to single gluon exchange,
$\sigma_1^{(4)}$ (Fig. \ref{ladder2fig}b) is given from diagrams with 
one gluon on one side of the cut and three on the other. In this case,
the transfered momentum and colour is given from that of the
single gluon. In all the three cases, we will assume that each 
individual gluon, identified by its origin from the pomerons, has 
the same momentum distribution as determined in the diffractive case, 
from the triple discontinuity calculation. In this 
way, we can calculate the momentum transfer also for the two 
non-diffractive contributions. The normalisation of each 
contribution is determined from the AGK cutting rules (and the
diffractive calculation), so that the relative size of the inclusive
cross-sections will satisfy Eq.~(\ref{agk}).

The applicability of the AGK rules in QCD was discussed for the case of
quark-quark scattering in the report article of Gribov, Levin and
Ryskin \cite{GLR} and the case of photon-proton scattering was treated more
recently, in a article by Bartels and Ryskin \cite{BRAGK}.
This latter treatment was based on contributions from triple 
discontinuities, as for the diffractive 
cross-section discussed  above, but for arbitrary colour states. 

In the present model we are assuming that there are further important
contributions, apart from the triple discontinuities, which would
arise from the real parts of the relevant
amplitudes\footnote{Note that we here in general use the 
word ``amplitude'' for the diagram on either side of the
final-state cut.}.  We will not here attempt to calculate these 
sub-leading contributions from first principles, but the usage of the
AGK cutting rules suggest that these are necessary in our treatment. 

In the Appendix, we show, for the symmetric colour exchange part 
of $\sigma_2$, that the triple discontinuity result completely 
agrees with the AGK rules as they are used here.
We are thus assuming that $\sigma_2$ has important
contributions from the real part of the relevant amplitudes, but only in
the case that the two exchanged gluons are in a anti-symmetric colour
state.

We end this section with a more intuitive discussion
of the above results, in particular, the interpretation of the higher
order correction $\sigma_1^{(4)}$ and its AGK factor relative to
$\sigma_2$. Consider a model, e.g. the BFKL formalism, which 
approximates the cross-section with the single gluon
exchange cross-section, $\sigma_1^{(2)}$. The
inclusive cross-section can, in this approximation, be described as
the sum of all possibilities
to exchange a single gluon between the quark pair and the
proton. However, the possibility of the simultaneous exchange of a
second gluon is not excluded, but it goes beyond the accuracy of the
calculation. In the leading BFKL calculation, the correction is formally
small, since it
has two extra factors of the strong coupling but no new factors of 
${\rm log}(s)$. This approximation leads to an overestimate of the 
total cross-section because, roughly speaking, events with double gluon 
exchange contribute twice. On the other hand, the 
one gluon exchange approximation underestimates the momentum transfer 
to the quark pair, simply because two exchanged gluons in general carry
more momentum than one. When double gluon exchange corrections are
included, as in the present model, we have two separate kinds of
events: We have a new kind of events with two exchanged gluons, 
with the cross-section $\sigma_2$, and we have the single gluon 
exchange events with a modified cross-section
$\sigma_1=\sigma_1^{(2)}+\sigma_1^{(4)}$. The role of $\sigma_1^{(4)}$
is to correct for the double counting (in $\sigma_1^{(2)}$) of the 
double gluon exchange events. Therefore, one can intuitively expect
the AGK relation $\sigma_1^{(4)}=-2\sigma_2$ to be valid.

There are two important implications from this. First, that the total
higher order correction is negative. We know this because, $\sigma_2$
corresponds to the probability of a certain process, and must be
positive, while the total correction, 
$\sigma^{(4)}=\sigma_1^{(4)}+\sigma_2=-\sigma_2$ is negative. The
other implication, that we can expect and will become more
explicit in the treatment below, is that the differential
cross-section at a certain finite momentum transfer is safe towards
the exchange of a very soft gluon. If one of the two exchanged gluons
is very soft,  $\sigma_2$ will get a contribution at a momentum
transfer which is close to the momentum of the harder gluon. The
correction $\sigma_1^{(4)}$ on the other will get half of its
contribution at the softer momentum and half of it at the higher
one. At the larger momentum transfer, the two contributions will, in
the limit, cancel exactly, while there will be an extra contribution, 
which we usually do not have good control over, around the point of zero
momentum transfer. In a perturbative treatment, it is therefore better
to study corrections to the differential cross-section instead of the
inclusive.

\section{The coupling of two gluons to the photon}
\label{logs}
Let us start with a more detailed discussion of the amplitude which is
relevant for the exchange of two gluons 
in the photon-proton scattering process. On the upper side, we have a
virtual photon which breaks up into a quark-anti quark pair. On the lower side
we have the incoming proton. Two gluons are exchanged between the proton and
the quark pair, either as two separate exchanged gluons or they can combine
into one gluon which is absorbed by the quark-pair. The proton transforms 
into a hadronic state with high
mass, in the case of non-diffractive DIS, and a lower mass state in case of 
pomeron exchange. 

The two gluons can be in either of the colour states of a singlet,
an anti-symmetric octet or a symmetric octet. Higher representations are
not possible for the gluons in this process since the outgoing quark pair
lies in the $3\otimes\bar{3}=1\oplus 8$ colour space. We will now consider 
the different colour
states separately and describe them in the frame work of the leading logs
expansion (LLs). Denote the amplitude for singlet exchange
${\cal A}_1$, for anti-symmetric octet exchange ${\cal A}_{8a}$ and for the 
symmetric octet ${\cal A}_{8s}$. The leading part of 
${\cal A}_{8a}$ is known. It factorises 
into the amplitude for one-gluon exchange and, to ${\cal O}(\alpha_s^2)$,
a factor $\omega(t){\rm log}(s)$ where
$\omega$ is the trajectory function of a reggeised gluon and $t$ is the
momentum transfer. In the partonic language, it
corresponds to the exchange of a single gluon, and the factor
$\omega(t){\rm log}(s)$ corresponds to a virtual emission. In our case we will be
interested in contributions that belong to the next order of the LLs
expansion, that is, they lack any factors of ${\rm log}(s)$, but contain two
factors of the strong coupling $\alpha_s$.

In the following, it will be useful to separate the amplitudes into their real
and imaginary parts. The reason is that the imaginary parts are (for
asymptotic s) easily calculable through the Cutkoski rule. We can then use
the AGK cutting rules to also determine some contributions, due to the real 
parts of the amplitudes, which are important in the model.
In Regge theory, it is expected that both of the 
amplitudes ${\cal A}_1$ and ${\cal A}_{8s}$ correspond to the exchange of
even signature
reggeons. The leading part of a reggeon with even signature is expected to be 
purely imaginary, thus we can assume ${\cal A}_1={\rm iIm}{\cal A}_1$ and 
${\cal A}_{8s}={\rm iIm}{\cal A}_{8s}$. The leading part of ${\cal A}_{8a}$ is 
real (as mentioned above) while
the next-to leading contribution has both a real and an imaginary part. In our
case, when the two gluon exchange amplitude is multiplied with its conjugate,
the contributions ${\cal A}_1{\cal A}_1^\dagger$, 
${\cal A}_{8s}{\cal A}_{8s}^\dagger$ and 
${\rm Im}{\cal A}_{8a}{\rm Im}{\cal A}_{8a}^\dagger$ can be
calculated in the asymptotic limit, using the
Cutkosky relation, while the relevant contributions from 
${\rm Re}{\cal A}_{8a}^{\rm (NLLs)}{\rm Re}
{\cal A}_{8a}^{\dagger\rm (NLLs)}$ can
be related to the others by the AGK cutting rules
\footnote{This does, in no way, lead to a full determination of 
${\rm Re}{\cal A}_{8a}^{\rm (NLLs)}$. It will give determination of the part of  
${\rm Re}{\cal A}_{8a}^{\rm (NLLs)}$, which contains one gluon from each
pomeron, so that  
${\rm Re}{\cal A}_{8a}^{\rm (NLLs)}{\rm Re}{\cal A}_{8a}^{\dagger\rm (NLLs)}$
grows with energy, as fast as the other terms.
We will in this way, not learn anything of e.g. the vertex 
corrections for the reggeised gluon.}.
\subsection{The four-gluon impact factor}
\begin{figure}
(a)
\psfig{figure=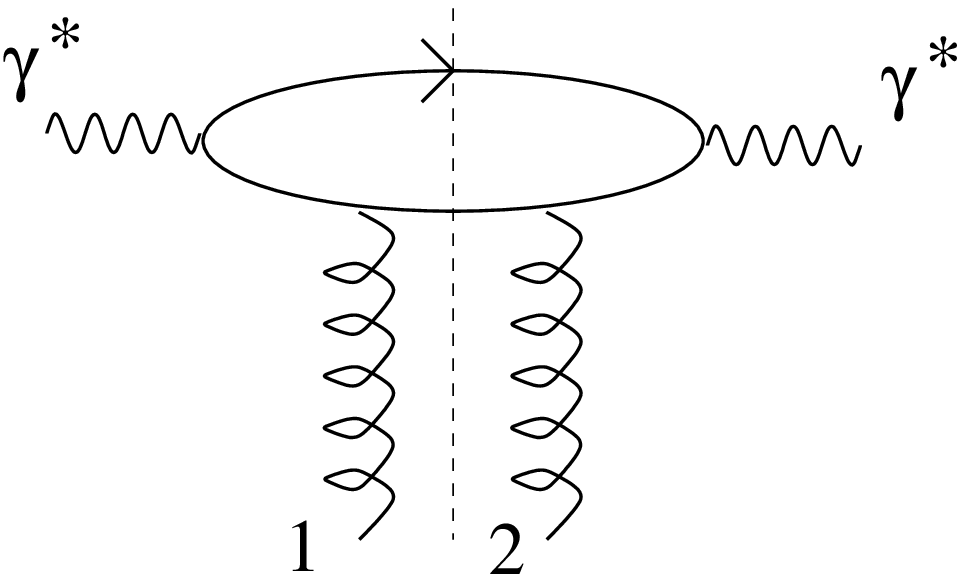,width=6cm}
\\
\vskip 1cm
(b)
\psfig{figure=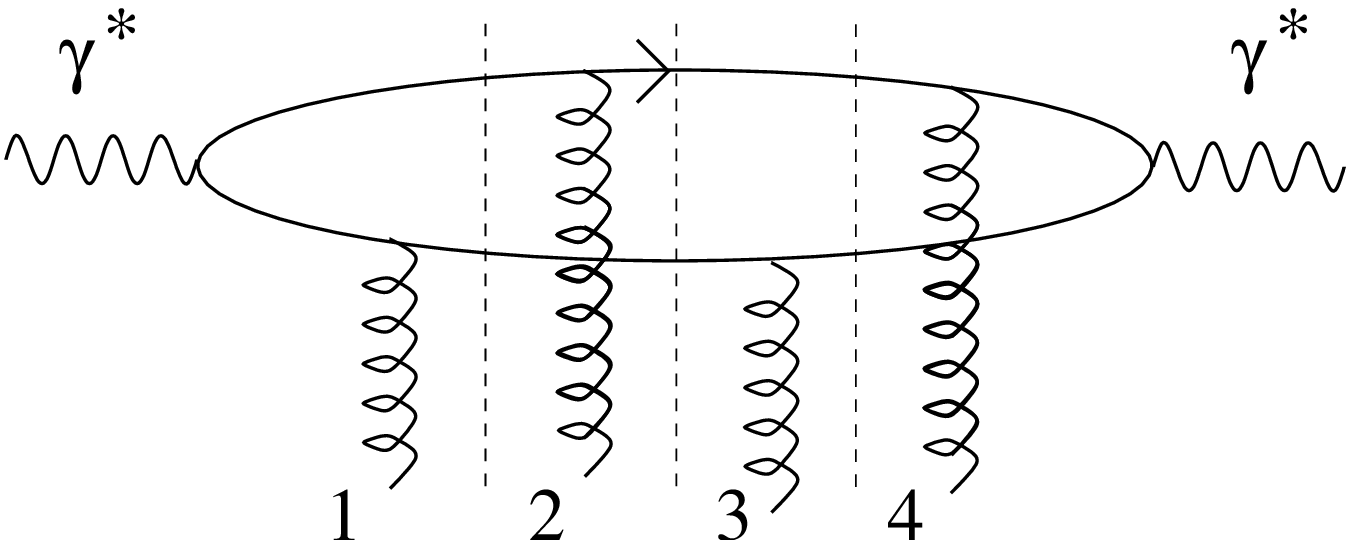,width=9cm}
\caption{(a) One of the four Feynman diagrams that contribute to the two-gluon
  impact factor of the photon, $D_{(2;0)}({\bf k}_1,{\bf k}_2)$. 
  (b) One of the 16 contributions to
  the impact factor $D_{(4;0)}({\bf k}_1,{\bf k}_2,{\bf k}_3,{\bf k}_4)$. 
  The triple-discontinuity is relevant for
  the diffractive cross-section, since the amplitude for pomeron exchange is
  purely imaginary, to leading order. Note that cut particles are on 
  mass-shell, while the gluons (and the photon) can be virtual.}
\label{imp24}
\end{figure}

\begin{figure}
\psfig{figure=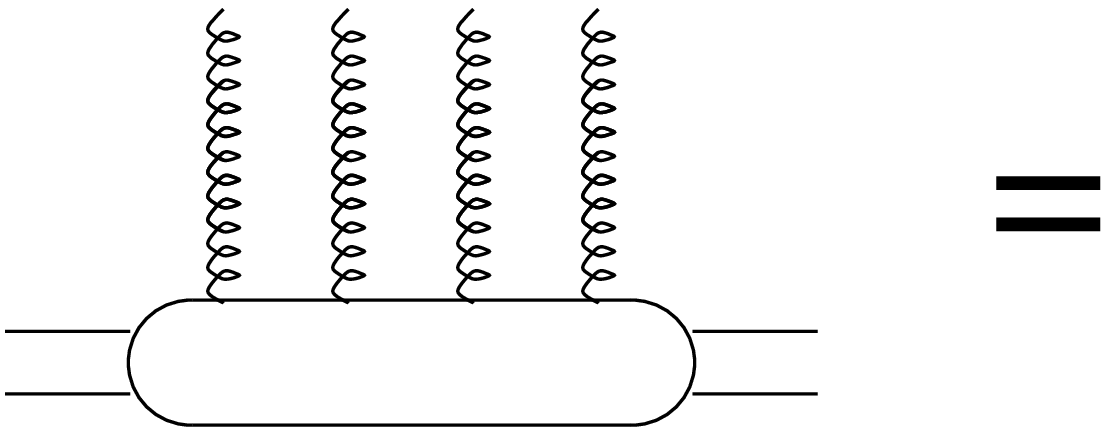,width=5cm}
\\
\vskip 1cm
\psfig{figure=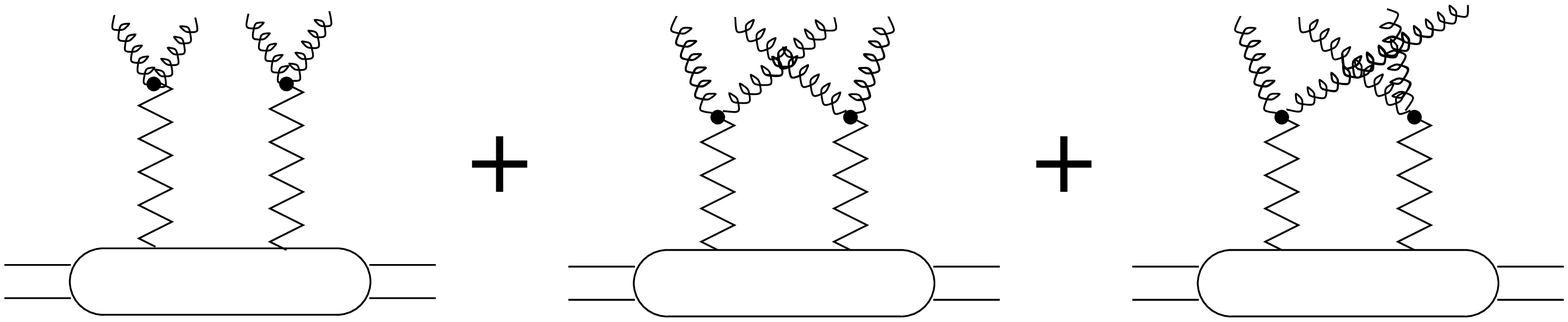,width=14cm}
\caption{The assumed form for the four-gluon impact factor of the proton. Only
  the first term contributes to the diffractive cross-section, $\sigma_0$. The
  other two terms are needed in the consistency check for the AGK rules 
  (In the Appendix).}
\label{proto}
\end{figure}

For the diffractive cross-section, denoted $\sigma_0$ above, we will
use the result of \cite{D4}, where the four-gluon impact factor of the
photon is calculated in the asymptotic limit. This impact factor is given by
multiply cut Feynman diagrams where the four gluons couple to a photon via
a quark loop (See Fig. \ref{imp24}b). $\sigma_0$ is then given by convoluting 
this impact factor with the part of the proton's impact factor which
contributes to pomeron exchange 
(That is, with one pomeron-proton vertex on each side of the final state
cut, the first term of the second row in Fig. \ref{proto}). The proton's 
impact factor is discussed more detailed in the next
section. There, we will relate it to the impact factor that is used in single
pomeron exchange and to the phenomenology of high energy elastic and 
diffractive scatterings.

We will now explain why the multiply cut photon impact factor is relevant for the
calculation of $\sigma_0$. The main argument is that the leading part of the 
amplitude for pomeron exchange is purely imaginary. According to the Cutkoski
rule, it is then given by its s-discontinuity times a factor $i/2$. For the
${\cal O}(\alpha_s^2)$ amplitude that we are considering, taking the
s-discontinuity means that we
have to take into account all planar diagrams (where the gluons do not cross or
interact) and put a cut in between the two gluons. The practical meaning of
the cut is that all particles  that cross it are on mass shell. When the 
amplitude is multiplied
with its conjugate, we will end up with planar diagrams with four-gluons
exchanged between the proton and the quark loop and with cutting lines in
between all the gluons. The sum of the upper parts of these diagrams will then
directly correspond to the multiply cut photon impact factor defined in 
\cite{D4}.

Before describing the explicit expression for the photonic four-gluon
impact factor, it is useful to discuss the corresponding one with only two 
gluons (Fig. \ref{imp24}a).
This impact factor is given by planar diagrams where two gluons couple to the
quark loop in all possible ways and with a cut in between the gluons.
It has the form of a colour tensor, with index $a_1$ and $a_2$ for
gluon one and two respectively, and is a function of the transverse momenta
(In Breit frame), ${\bf k}_1$ and ${\bf k}_2$, of the gluons. It is given by
\begin{equation}
\label{d2nf}
D_{(2;0)}^{a_1a_2}({\bf k}_1,{\bf k}_2)=
g^2\delta_{a_1a_2}\frac{1}{2}
\left[
D_{(2;0)}({\bf k}_1)+D_{(2;0)}({\bf k}_2)-D_{(2;0)}({\bf k}_1+{\bf k}_2)
\right]
\end{equation}
where $D_{(2;0)}({\bf k})$ denotes the forward impact factor which, after
the loop integration, has the analytic form \cite{D2}
\begin{equation}
\label{d2an}
D_{(2;0)}({\bf k})=\frac{1}{2(2\pi)^2}\sum_f e_f^2 
\int_0^1d\alpha
\int_0^1dy
\frac{[1-2\alpha(1-\alpha)][1-2y(1-y)]k^2}
{\alpha(1-\alpha)Q^2+y(1-y)k^2}
\end{equation}
Here, $Q^2$ is the virtuality of the photon, $\alpha$ is the energy fraction
of one of the quarks and $y$ is a Feynman parameter introduced in order to do
the transverse momentum integration. The sum is over quark flavours. This
expression is valid for transverse polarisations of the photon and for
massless quarks. For simplicity, we will disregard longitudinal polarisation
and quark masses. This impact factor has the important property 
that it goes linearly to zero whenever one of its arguments goes to zero. The
interpretation is that when the gluon becomes soft, it has a long wavelength
and can not resolve the quark anti-quark pair. Since the quark pair is in a
singlet system, the gluon does not interact.

Later on, it will be useful to write the two-gluon impact factor of the
proton, in the form
\begin{equation}
\label{phi2}
\phi_2^{a_1a_2}({\bf k}_1,{\bf k}_2)=
g^2\frac{N}{N^2-1}\delta_{a_1a_2}f_2({\bf k}_1,{\bf k}_2),
\end{equation}
where $N$ is the number of colours ($N=3$). With this definition, we can
identify the function $f_2$ in the forward limit, with the gluon distribution
in the proton. The cross-section for $\gamma^*$-proton scattering, to 
${\cal O}(\alpha_s^2)$ is given by
$$
\sigma_1^{(2)}=D_2 \otimes \phi_2.
$$
The sub-script specifies that this is a contribution to the one gluon
exchange process and the (2) denotes the order in $\alpha_s$. 
The convolution procedure is defined by summation over colour indices and
integration over the transverse momentum loop, with the measure\footnote{Note 
that the convolution procedure has the dimension of cross-section.}
$$
d{\cal K}_2\equiv
\frac{d^2{\bf k}_1}{(2\pi)^2}
\frac{d^2{\bf k}_2}{(2\pi)^2}
\frac{(2\pi)^2\delta({\bf k}_1+{\bf k}_2)}{k_1^2k_2^2}.
$$
For the momentum transfer, $p_\perp$, we can write the differential cross
section
\begin{equation}
\label{12diff}
\frac{d\sigma_1^{(2)}}{dp_\perp^2}=
\pi g^4N
\int d{\cal K}_2\delta ({\bf k}_1-{\bf p}_\perp)
f_2({\bf k}_1,{\bf k}_2)D_2({\bf k}_1,{\bf k}_2)
\end{equation}

We can now write the explicit form of the four-gluon impact factor \cite{D4}
\begin{eqnarray}
\label{d401}
D_{(4;0)}^{a_1a_2a_3a_4}({\bf k}_1,{\bf k}_2,{\bf k}_3,{\bf k}_4)&=&  \\
g^4d^{a_1a_2a_3a_4}D_{(4;0)}^A({\bf k}_1,{\bf k}_2,{\bf k}_3,{\bf k}_4)&+&
g^4d^{a_2a_1a_3a_4}D_{(4;0)}^B({\bf k}_1,{\bf k}_2,{\bf k}_3,{\bf k}_4) \nonumber
\end{eqnarray}
with
\begin{eqnarray}
\label{d402}
D_{(4;0)}^A({\bf k}_1,{\bf k}_2,{\bf k}_3,{\bf k}_4)&=&
D_{(2;0)}({\bf k}_1+{\bf k}_2+{\bf k}_3,{\bf k}_4)
+D_{(2;0)}({\bf k}_1,{\bf k}_2+{\bf k}_3+{\bf k}_4) \nonumber \\
&-&D_{(2;0)}({\bf k}_1+{\bf k}_4,{\bf k}_2+{\bf k}_3), \\
D_{(4;0)}^B({\bf k}_1,{\bf k}_2,{\bf k}_3,{\bf k}_4)&=&
D_{(2;0)}({\bf k}_1+{\bf k}_3+{\bf k}_4,{\bf k}_2)
+D_{(2;0)}({\bf k}_1+{\bf k}_2+{\bf k}_4,{\bf k}_3) \nonumber \\
&-&D_{(2;0)}({\bf k}_1+{\bf k}_2,{\bf k}_3+{\bf k}_4)
-D_{(2;0)}({\bf k}_1+{\bf k}_3,{\bf k}_2+{\bf k}_4). \nonumber
\end{eqnarray}
Here, $g$ is the strength of the strong coupling and $d$ is a colour tensor 
defined by $d^{abcd}={\rm tr}[t^at^bt^ct^d]+{\rm tr}[t^dt^ct^bt^a]$, where $t^a$
denote the SU(3) generators. It will be useful for the following, to separate
$D_{(4;0)}$ into two parts. One part, with the gluons one and two (and 
consequently gluons 3 and 4) in a symmetric colour state, and one part for the
anti-symmetric colour state. This separation is unique ($(+)$ for symmetric
and $(-)$ for anti-symmetric).
\begin{equation}
\label{separation}
D_{(4;0)}^{a_1a_2a_3a_4}({\bf k}_1,{\bf k}_2,{\bf k}_3,{\bf k}_4)=
g^4d_+^{a_1a_2a_3a_4}D_{(4;0)}^{(+)}({\bf k}_1,{\bf k}_2,{\bf k}_3,{\bf k}_4)+
g^4d_-^{a_1a_2a_3a_4}D_{(4;0)}^{(-)}({\bf k}_1,{\bf k}_2,{\bf k}_3,{\bf k}_4)
\end{equation}
with 
\begin{eqnarray}
\label{symmdef}
d_+^{a_1a_2a_3a_4}&\equiv&\left(d^{a_1a_2a_3a_4}+d^{a_2a_1a_3a_4}\right)/2 \\
d_-^{a_1a_2a_3a_4}&\equiv&\left(d^{a_1a_2a_3a_4}-d^{a_2a_1a_3a_4}\right)/2 
\nonumber\\
D_{(4;0)}^{(+)}({\bf k}_1,{\bf k}_2,{\bf k}_3,{\bf k}_4)&\equiv&
D_{(4;0)}^A({\bf k}_1,{\bf k}_2,{\bf k}_3,{\bf k}_4)+
D_{(4;0)}^B({\bf k}_1,{\bf k}_2,{\bf k}_3,{\bf k}_4) \nonumber\\
D_{(4;0)}^{(-)}({\bf k}_1,{\bf k}_2,{\bf k}_3,{\bf k}_4)&\equiv&
D_{(4;0)}^A({\bf k}_1,{\bf k}_2,{\bf k}_3,{\bf k}_4)-
D_{(4;0)}^B({\bf k}_1,{\bf k}_2,{\bf k}_3,{\bf k}_4) \nonumber
\end{eqnarray}

The impact factor $D_{(4;0)}$ vanishes linearly whenever
$k_1$ or $k_4$ becomes soft. This is, in general not the case for $k_2$ and
$k_3$. This can be interpreted as that the gluons 1 and 4 interact with a
colour less quark pair, while gluons 2 and 3 interact with the quarks in an 
octet state. One can however check that the symmetric part, $D_{(4;0)}^{(+)}$,
vanishes with any of its argument. Further, it is invariant under the
permutation of the momenta of the gluons. We know this because it can be 
written as the sum of permutations of 
$D_{(2;0)}({\bf k}_1,{\bf k}_2+{\bf k}_3+{\bf k}_4)$ and
$-D_{(2;0)}({\bf k}_1+{\bf k}_2,{\bf k}_3+{\bf k}_4)$.
(Note that $D_{(2;0)}({\bf k}_1,{\bf k}_2)=D_{(2;0)}({\bf k}_2,{\bf k}_1)$.)

The diffractive cross-section $\sigma_0$, which will be determined in the
following, will have all of its contribution given from $D_{(4;0)}^{(+)}$. In
our treatment, this will be the case also for the non-diffractive
$\sigma_2$ and $\sigma_1^{(4)}$, since these are determined from $\sigma_0$
and the AGK rules. This means that we are assuming that the non-diffractive
cross-sections have additional contributions other than from the multiply cut
impact factor. This could, for example be from Feynman diagrams with triple
gluon vertices, where two gluons merge into one before they are absorbed by
the quark pair. One can speculate that such contributions are important,
because they allow the "second" exchanged gluon to, in all possible ways,
interact with the system of the quark, anti-quark and the "first" gluon, which
in total is in the colour singlet state.

\subsection{The diffractive contribution}
For the part of the proton impact factor, denoted $\phi_{4({\cal P})}$,
 which is used in the calculation of $\sigma_0$, we assume the form
\begin{equation}
\label{p0}
\phi_{4({\cal P})}^{a_1a_2a_3a_4}({\bf k}_1,{\bf k}_2,{\bf k}_3,{\bf k}_4)=
g^4\frac{N^2}{(N^2-1)^2}\delta_{a_1a_2}\delta_{a_3a_4}
f_4({\bf k}_1,{\bf k}_2;{\bf k}_3,{\bf k}_4)
\end{equation}
The Kronecker deltas assure colour singlet exchange, and the function $f$
will be determined in the next section. For now, it is sufficient to mention
that it is dimension less, it goes linearly to zero whenever one of its 
argument goes to zero, and it has the following symmetry properties:
\begin{eqnarray}
f_4({\bf k}_1,{\bf k}_2;{\bf k}_3,{\bf k}_4)&=&
f_4({\bf k}_2,{\bf k}_1;{\bf k}_3,{\bf k}_4) \nonumber \\
&=&f_4({\bf k}_1,{\bf k}_2;{\bf k}_4,{\bf k}_3) \nonumber \\
&=&f_4({\bf k}_3,{\bf k}_4;{\bf k}_1,{\bf k}_2). \nonumber
\end{eqnarray}

The cross-section $\sigma_0$, for pomeron exchange, is now given by the
convolution of the two impact factors, $D_{(4;0)}$ and $\phi_{4({\cal P})}$,
$$
\sigma_0=\frac{1}{4}D_{(4;0)}\otimes\phi_{4({\cal P})}
$$ 
where the factor of one fourth is because of the
double use of the Cutkoski rules.

Later on, in section \ref{num}, we will be interested in the momentum transfer 
between the proton and the quark pair. Therefore, we write in more
detail, the differential cross-section for a transverse momentum exchange
$p_\perp$ 
\begin{eqnarray}
\label{s0}
\frac{d\sigma_0}{dp_\perp^2}&=&
\frac{1}{4}\cdot \pi g^8\frac{N}{2} \\
&\times&
\int
d{\cal K}_4\delta({\bf k}_1+{\bf k}_2-{\bf p}_\perp)
f_4({\bf k}_1,{\bf k}_2;{\bf k}_3,{\bf k}_4)
D_{(4;0)}^{(+)}({\bf k}_1,{\bf k}_2,{\bf k}_3,{\bf k}_4) \nonumber \\
d{\cal K}_4&\equiv&
\frac{d^2{\bf k}_1}{(2\pi)^2}
\frac{d^2{\bf k}_2}{(2\pi)^2}
\frac{d^2{\bf k}_3}{(2\pi)^2}
\frac{d^2{\bf k}_4}{(2\pi)^2}
\frac{(2\pi)^2\delta({\bf k}_1+{\bf k}_2+{\bf k}_3+{\bf k}_4)}{k_1^2k_2^2k_3^2k_4^2}
 \nonumber
\end{eqnarray}
For the colour part
of the problem, we have used the result (with $d_+$ defined in
eq. (\ref{symmdef})). 
$$
\delta_{ab}\delta_{cd}d_+^{abcd}=2C_f^2N
$$
Where $C_f=(N^2-1)/2N$ is the colour factor for gluon emission from a quark.

\subsection{Non-diffractive contributions}
We will now make use of the AGK rules to relate Eq. (\ref{s0}) to the 
processes of non-diffractive double gluon exchange and single gluon exchange. 
It is then important to note that ${\bf k}_1$ and ${\bf k}_2$ are the internal
momenta of the pomeron, up to now, on one side of the final-state cut, 
and ${\bf k}_3$, ${\bf k}_4$ are 
the momenta of the pomeron on the other side. The AGK cutting rules relate 
the contribution $\tilde{\sigma}_0$ with no cut pomerons to the contributions 
$\tilde{\sigma}_1$ of one cut and $\tilde{\sigma}_2$ of two cut pomerons. 
We will now identify $\tilde{\sigma}_0$ with the diffractive cross-section
$\sigma_0$, which we have considered to now, and $\tilde{\sigma}_2$ and 
$\tilde{\sigma}_1$, with the cross-section for non-diffractive double gluon
exchange $\sigma_2$, and the higher order correction to single gluon exchange
$\sigma^{(4)}_1$, respectively. We assume further that the relation
$\sigma_2:\sigma^{(4)}_1:\sigma_0=2:-4:1$ is valid on a more exclusive level
with fixed values of the transverse momenta $k_i$. 

For clarity in the following two
equations, we will change notations: ${\bf k}_1={\bf k}_{a1}$, 
${\bf k}_2={\bf k}_{{\rm a}2}$ for the gluons that emerge from the pomeron a,
and ${\bf k}_3={\bf k}_{{\rm b}1}$, ${\bf k}_4={\bf k}_{{\rm b}2}$, for the 
others. In the case of two cut pomerons, the momentum transfer 
(The sum of the momenta on either side of the cut) can be written as either 
${\bf p}_\perp={\bf k}_{{\rm a}1}+{\bf k}_{{\rm b}1}$ or 
${\bf p}_\perp={\bf k}_{{\rm a}1}+{\bf k}_{{\rm b}2}$. 
Because of symmetry properties of the
impact factors, the two possibilities give the same result, and choosing the
first one we get

\begin{eqnarray}
\label{s2}
\frac{d\sigma_2}{dp_\perp^2}&=&
\frac{1}{2}\cdot \pi g^8\frac{N}{2} \\
&\times&
\int
d{\cal K}_4\delta({\bf k}_{{\rm a}1}+{\bf k}_{{\rm b}1}-{\bf p}_\perp)
f_4({\bf k}_{{\rm a}1},{\bf k}_{{\rm a}2};
{\bf k}_{{\rm b}1},{\bf k}_{{\rm b}2})
D_{(4;0)}^{(+)}({\bf k}_{{\rm a}1},{\bf k}_{{\rm a}2},
{\bf k}_{{\rm b}1},{\bf k}_{{\rm b}2}) \nonumber
\end{eqnarray}

With $d{\cal K}_4$ defined in Eq. (\ref{s0}), $f_4$ defined in Eq. (\ref{p0})
and  $D_{(4;0)}^{(+)}$ defined in Eqs. (\ref{symmdef}) and (\ref{d402}).
For the case of one cut pomeron, we have any of the possibilities 
${\bf p}_\perp={\bf k}_i$
but again the choice is irrelevant\footnote{Note that the number of
  independent choices for $p_\perp$, coincides with the AGK weights.} 
 With ${\bf p}_\perp={\bf k}_{{\rm a}1}$ we get

\begin{eqnarray}
\label{s1}
\frac{d\sigma_1^{(4)}}{dp_\perp^2}&=&
(-1)\cdot \pi g^8\frac{N}{2} \\
&\times&
\int
d{\cal K}_4\delta({\bf k}_{{\rm a}1}-{\bf p}_\perp)
f_4({\bf k}_{{\rm a}1},{\bf k}_{{\rm a}2};
{\bf k}_{{\rm b}1},{\bf k}_{{\rm b}2})
D_{(4;0)}^{(+)}({\bf k}_{{\rm a}1},{\bf k}_{{\rm a}2},
{\bf k}_{{\rm b}1},{\bf k}_{{\rm b}2}) \nonumber
\end{eqnarray}

It is important to note that we do not consider all exchanges which are 
in the colour singlet state, as pomeron exchanges. Above, the pomeron was
described by two gluons in colour singlet. For the contributions with two cut
pomerons, we have, on each side of the cut, one gluon from each pomeron. These
two gluons could very well also be in a colour singlet state, but we would not 
consider the process to be pomeron exchange. (This contribution is however 
colour suppressed with a factor $1/(N^2-1)$.) It is not certain that the 
(partonic) process of the exchange of two gluons in a singlet state, as 
opposed to pomeron exchange, should lead to a rapidity gap in the final-state 
distribution of hadrons, since that would correspond to
(non-perturbative) colour screening, in the hadronisation process 
(See e.g. \cite{gunnar}). We will leave this question open, but refer to the 
contribution of $\sigma_2$ as non-diffractive.

We will now summarise the main assumptions that was needed to derive the main
result of this section, that is, Eqs. (\ref{s2}) and
(\ref{s1}). We started by considering the process of pomeron exchange. There,
we assumed that the amplitude is imaginary to leading order and we could 
therefore use the four-gluon impact factor calculated in \cite{D4}. Further 
on, we made some assumptions on the form of the proton impact factor 
(The part of it which corresponds to pomeron exchange), including some
symmetry properties. Finally we assumed that the AGK cutting 
rules are valid so that we could relate the process of pomeron 
exchange to the processes of double gluon exchange and a sub-leading
correction to single gluon exchange. The
AGK rules where assumed to be valid for fixed values of the internal momenta 
(the momenta of the two gluons) in each pomeron, while originally \cite{AGK}, 
they were stated as a relation between the inclusive cross-sections.

In section \ref{num}, we will use the Eqs. (\ref{s2}) and (\ref{s1}) to make
some estimates of observable effects of double gluon exchange. In the next
section, the impact factor of the proton, is discussed and determined on a
more quantitative level.

\section{The proton impact factor}
\label{prota}
We will start with a discussion of the two-gluon impact factor, $\phi_2$, of 
the proton. This is related to the conventional unintegrated structure 
function or the parton distribution of the proton, and there are a large 
number of calculations and measurements available for its determination. It 
would, however, be inconsistent to make a direct use of the available parton 
distributions, since these have, in the normal case, 
been fitted to data, without considering effects from multiple interaction.
Instead, we make a simple anzats for $\phi_2$, by using the measured parton
distributions at large momentum fractions, $x_p\sim 0.1$, where corrections
from double gluon exchange should be small, and by assuming an 
exponential growth towards lower $x_p$-values. 

In the second part of this section, we will determine the four-gluon impact
factor, $\phi_4$, from the square of the parton distribution, times a normalization
factor which is determined from two separate measurements in proton-anti-proton
scattering. The uncertainty in $\phi_2$, and consequently in $\phi_4$
will lead to one of the main uncertainties in the final result.
In a more careful treatment, it is possible to
determine these functions more accurately than will be done here. We will
instead try to keep things as simple as possible, with the aim of just an
order of magnitude estimate in the final results. 

\subsection{The two-gluon impact factor}
With our way of defining the two-gluon impact factor, in Eq. (\ref{phi2}), 
the function $f_2({\bf k}_1,{\bf k}_2)$ is, in the forwards limit, given
by the number of gluons, $n(k)$, inside the proton, that an exchanged gluon,
with transverse momentum $k$ can resolve.
\begin{equation}
\label{f2n}
f_2({\bf k},-{\bf k})=n(k)
\end{equation}

The exact meaning of "resolve", in this case, is that we require the
transverse momentum of the exchanged gluon to be smaller than half of the
invariant energy of one of the quarks, from the photon side, together
with the "resolved"  gluon in the proton, $k^2\!<\!\hat{s}/4$. 
This leads to a value,
$\hat{x}(k^2)$, for the smallest possible momentum fraction, $x_p$, that a 
resolved gluon can have. Apart from $k^2$, this limit ($\hat{x}$) depends on the
proton-photon CMS energy ($W$) and the momentum fraction, $\alpha$ of the 
participating quark from the photon side
$$
x_p>\frac{4}{\alpha}\frac{k^2}{W^2}\equiv \hat{x}.
$$
In any case, we will always require that $x_p\!>\! x_B/\alpha$. We will not try
to determine $\alpha$ from event to event, but instead, in this case, use the 
typical value $\alpha =0.5$.

\begin{figure}
\begin{picture}(0,245)
\large
\put(0,230){$xg(x)$}
\normalsize
\put(28,206){30}
\put(28,146){20}
\put(28,85){10}
\put(34,30){0}
\put(16,10){\psfig{figure=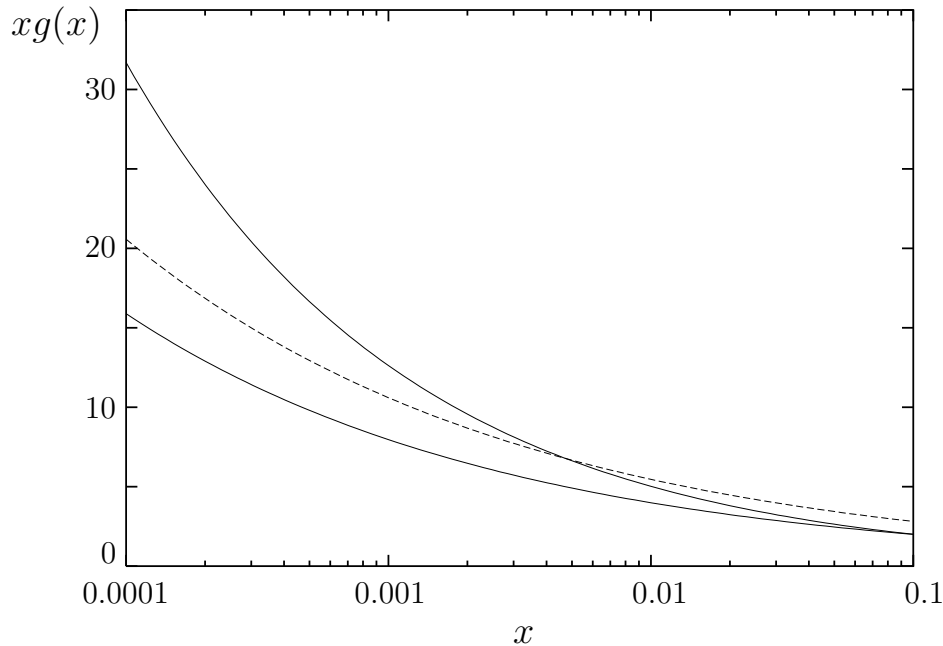,width=12cm}}
\large
\put(190,0){$x$}
\normalsize
\put(27,15){0.0001}
\put(130,15){0.001}
\put(235,15){0.01}
\put(338,15){0.1}
\end{picture}
\caption{The assumed gluon distribution, $xg(x)$, for $\lambda=0.3$ (lower
solid curve) and $\lambda=0.4$ (upper solid curve). The dashed curve is the
gluon distribution of the GWB model.}
\label{xg}
\end{figure}

To determine the number of resolved gluons, we will assume a
constant contribution of $n=2$ from the region $0.1\!<\!x_p\!<\!1$. We will 
further assume a density of two gluons per unit of ${\rm log}(x_p)$ 
at the point $x_p=0.1$ and from there, that it grows like $x_p^{-\lambda}$.
For the value of the exponent, we will consider the two cases $\lambda=0.3$
and $\lambda=0.4$, respectively. (The gluon distributions, 
$xg_\lambda(x)$, are plotted in Fig. \ref{xg}.)
This results in the following expression for the number of
gluons with $x_p\!>\!\hat{x}$
\begin{equation}
\label{hatn}
\hat{n}(\hat{x})=
2+
\frac{2}{\lambda}
\left(
\left(\frac{\hat{x}}{0.1}\right)^{-\lambda}-1
\right)
\theta(0.1-\hat{x})
\end{equation}
where, $\theta$ is the step-function.

Finally, we have to take into account coherence between the gluons inside the
proton. When the wavelength of the exchanged gluon is at the
order of the proton size, it cannot resolve the individual gluons and the
effective number of gluons will decrease. Therefore, we will assume that the 
effective number of gluons, goes to zero, linearly with $k^2$ and that the 
suppression starts around a scale $\mu^2$. This effect is taken into account 
by multiplying $\hat{n}(\hat{x})$ with a factor $k^2/(\mu^2+k^2)$. 
The suppression scale is set to $\mu^2=0.1$ ${\rm GeV}^2$ which is a typical
value for the momentum transfer in elastic scattering of hadrons (See the
discussion in the second part of this section).
This leads, finally, to
\begin{equation}
\label{nk}
n(k)=\frac{k^2}{\mu^2+k^2}\hat{n}
\left(\frac{4}{\alpha}\frac{k^2}{W^2}\right)
\end{equation}
with $\hat{n}$ defined in eq. (\ref{hatn}).

\begin{figure}
\begin{picture}(0,245)
\Large
\put(0,230){$n$}
\normalsize
\put(25,220){50}
\put(25,182){40}
\put(25,144){30}
\put(25,104){20}
\put(25,67){10}
\put(40,20){0}
\put(116,20){5}
\put(190,20){10}
\put(264,20){15}
\put(337,20){20}
\put(10,8){\psfig{figure=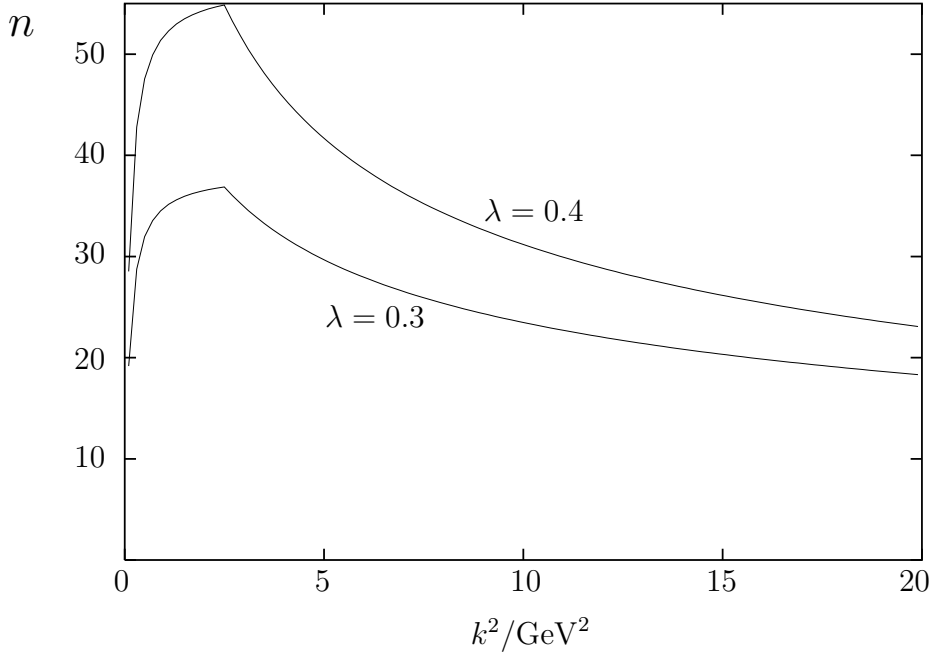,width=12cm}}
\put(175,0){$k^2/{\rm GeV}^2$}
\put(180,160){$\lambda=0.4$}
\put(120,120){$\lambda=0.3$}
\end{picture}
\caption{The number of resolved gluons $n(k^2)$, for $\lambda=0.3$ and 
$\lambda=0.4$, respectively.}
\label{pif}
\end{figure}

This function is plotted in Fig. \ref{pif}, again for $\lambda=0.3$ and
$\lambda=0.4$ and with $Q^2=10$~${\rm GeV}^2$ and $x_B=0.0001$. 
We note that the two $\lambda$-values result in almost a factor of two
difference in the low-$k^2$ region. This will be one of our main
uncertainties. As mentioned before, it would be
inconsistent to use the available measured parton distributions, since 
these have been fitted to data, without any account of multiple scattering.
Here, we have instead fixed the parton distribution at a large momentum
fraction $x_p=0.1$, where saturation effects should be small, and then
assumed a BFKL like $x_p$-dependence.

The turnover of $n(k^2)$ in Fig. \ref{pif}, occurs at the point where 
$\hat{x}=x_B/\alpha$. This happens at $k^2=x_BW^2/4=Q^2/4$.  

\subsubsection{A comparison with the Golec-Biernat W\"usthoff model}

As stated before, our parton distribution is poorly constrained,
because the available fits are in general made without any
concern to multiple interaction effects. There are however several
saturation models that have been fitted to the structure function,
and it is in some cases possible to extract the corresponding parton
distribution from these fits. We will use the model of Golec-Biernat
and W\"usthoff \cite{GBW1,GBW2}, which is perhaps the simplest 
one available and which describes the measured structure function 
very well, over a large range of $Q^2$, in particular down to the 
photo-production region.

Let us start with a minimal recollection of the GBW model. The
description of the DIS process is as follows, that the virtual photon
splits into a quark-anti-quark pair which then propagates through the
proton and exchanges one or several gluons. The $\gamma^*$-proton
cross-section is, in the GBW model given by
\begin{equation}
\sigma_{\pi}(x,Q^2)=\int d^2{\bf r}
\left|
\Psi_\pi({\bf r}) \right|^2
\hat{\sigma}(x,r^2)
\end{equation}

where $\Psi$ is the wave function of the photon (Integrated over the
energy fraction of the quarks), $\pi$ is the polarisation of the
photon and $\bf r$ is a vector in impact parameter space.
The dipole cross-section $\hat{\sigma}$ is, in this model, set to be
\begin{equation}
\label{dipole}
\hat{\sigma}(x,r^2)=\sigma_0
\left\{1-{\rm exp}\left(-\frac{r^2}{4R_0^2(x)}\right)\right\}
\end{equation} 
with 
\begin{equation}
R_0(x)=\frac{1}{Q_0}\left(\frac{x}{x_0}\right)^{\lambda/2}
\end{equation}
The parameters are $\sigma_0$, $x_0$ and $\lambda$, while $Q_0$ is set
to 1~GeV and is no real parameter.

A fit to HERA data resulted in the following values \cite{GBW1}:
$\sigma_0=23.03$~mb, $x_0=3.04\cdot 10^{-4}$ and $\lambda=0.288$.

The validity of the GBW model lies in that it is a simple model
which, with a few parameters, can describe the structure function over
a large range of $Q^2$. In fact, the authors refer to it as a model
which contains saturation, but they only briefly mention multiple
interactions as a possible explanation for saturation \cite{GBW2}.

It lies however close at hand to interpret the exponential in the
dipole cross-section (Eq.~\ref{dipole}) as the poissonian probability
of not having any interactions. (The probability of not having any
randomly scattered partons inside an area $r^2$, with the average 
density of partons $1/(4R_0^2)$.) The GBW model is thus a model 
which incorporates shadowing to all orders in the number of
exchanged gluons. An expansion of the exponent gives us the corrections
that appear when each higher order is taken into account.

A formula, which is useful for extracting the gluon distribution,
gives the following relation between the dipole cross-section and the
(gluonic) unintegrated structure function ${\cal F}_g(x,k^2)$
(Defined as a distribution in $k^2$, as in \cite{GBW2})
\begin{equation}
\hat{\sigma}(x,r)=\frac{4\pi\alpha_s}{3}
\int\frac{d^2{\bf k}}{k^2}
\left[1-{\rm e}^{{\rm i}{\bf r\cdot k}}\right]
{\cal F}_g(x,k^2).
\end{equation}

This relation can be inverted, but if we use $\hat{\sigma}$ as it
is, then ${\cal F}$ will not correspond to the gluon distribution.
That would be true only if we knew that the dipole absorbs
one and only one gluon in each collision. 

It is however possible to get the gluon distribution of the GBW model
by expanding the dipole cross-section in orders of exchanged gluons,
that is, expanding the exponent of $\hat{\sigma}$. The first term, then
would give us the structure function, if multiple interactions were
absent. This structure function would fit data poorly, 
especially in the low-$Q^2$ region, but it would instead directly 
corresponds to the gluon distribution. 

Using the first term in an expansion of $\hat{\sigma}$ we get
\begin{equation}
\sigma_0\frac{r^2}{4R_0^2(x)}=\frac{4\pi\alpha_s}{3}
\int\frac{d^2{\bf k}}{k^2}
\left[1-{\rm e}^{{\rm i}{\bf r\cdot k}}\right]
{\cal F}_g^1(x,k^2),
\end{equation}
where the superscript indicates that ${\cal F}_g^1$ is calculated 
from single gluon exchange. This equation is inverted to give
\begin{equation}
{\cal F}_g^1(x,k^2)=\frac{\sigma_0}{4R_0^2}\frac{3}{\pi^2\alpha_s}
\delta(k^2)
\end{equation}
where $\delta$ is Dirac's function. We thus find that the first
term in the expansion of $\hat{\sigma}$ gives a distribution peaked 
at zero transverse momentum. The full $\hat{\sigma}$, on the other
hand, gives a broader distribution, which is peaked at a finite
$k^2$. (When $xg(x,k^2)$ is integrated over
all $k^2$ we get however the same result in both cases.) 

For the gluon distribution, we finally get
\begin{equation}
xg(x,Q^2)\equiv\int^{Q^2}_0{\cal F}_g^1(x,k^2)dk^2=
\frac{\sigma_0}{4R_0^2(x)}\frac{3}{\pi^2\alpha_s}
\end{equation}

The gluon distribution that we find from the GBW model is thus
independent of $Q^2$. This could be expected, since there is no
transverse momentum evolution of the parton distribution incorporated
into the model. The GBW model thus fails to describe the structure
function in the high $Q^2$ limit, but the fit shows that
the strong $Q^2$ dependence in the lower region can be explained
as a saturation effect, and is not necessarily related to the usual
DGLAP scaling violation mechanism. 

Inserting the fitted parameters we get
\begin{equation}
xg(x)\approx\frac{0.436}{\alpha_s}x^{-\lambda}.
\end{equation}

This is plotted in Fig. \ref{xg} for $\alpha_s=0.3$. Note that $x$ in
the GBW model corresponds to $x_B$ of the e-p scattering. The momentum
fraction of the interacting gluons is not well defined in the model, 
but is in general somewhat larger than $x_B$. This correction would
increase the gluon distribution. We conclude that our assumed gluon
distributions, with a relatively large exponent and independent of
$Q^2$, are supported by data.

\subsection{The four-gluon impact factor}
Now that we have $\phi_2$ in the forward limit, we can continue with the 
discussion of how it can be used to determine the two-pomeron-proton form
factor, $\phi_{4({\cal P})}$. We will rely
on the picture that  $\phi_2$ corresponds to a pomeron-proton vertex, and 
$\phi_{4({\cal P})}$ is, roughly speaking, given by the square of $\phi_2$.
We will also use the fact that in high energy elastic and diffractive
scattering, one observes low values (well below 1 ${\rm GeV}^2$) of the typical 
momentum transfer. In this article, we will be interested in the momentum transfer
in DIS processes, and for $p_\perp^2$ values above 1 ${\rm GeV}^2$. With all the
transverse momenta $k_i$ larger than the typical pomeron momentum, we can
write the function $f_4({\bf k}_1,{\bf k}_2;{\bf k}_3,{\bf k}_4)$, 
defined in eq. (\ref{p0}), in the
following approximative way (assuming $\sum_i{\bf k}_i=0$)
\begin{eqnarray}
\label{fapp}
f_4({\bf p}_1+{\bf r},-{\bf p}_1;{\bf p}_2-{\bf r},-{\bf p}_2)&=&
f_2({\bf p}_1+{\bf r},-{\bf p}_1)f_2({\bf p}_2-{\bf r},-{\bf p}_2)P({\bf r})
\nonumber \\
&\approx&
n(p_1)n(p_2)P({\bf r})
\end{eqnarray}
Here, we have made the substitutions
\begin{eqnarray}
\label{parsub}
{\bf k}_1&=&{\bf p}_1+{\bf r} \nonumber \\
{\bf k}_2&=&-{\bf p}_1  \\
{\bf k}_3&=&{\bf p}_2-{\bf r} \nonumber \\
{\bf k}_4&=&-{\bf p}_2 \nonumber 
\end{eqnarray}
and the function $P({\bf r})$ carries all the dependence on the pomeron momentum 
and a possible normalisation constant. 
This function will have the effect of an overall normalization factor, in the
final result
\begin{equation}
\label{norm}
{\cal N}\equiv \int P({\bf r}) \frac{d^2{\bf r}}{(2\pi)^2}.
\end{equation}

For the function $P({\bf r})$ we assume the simple exponential form 
\begin{equation}
\label{Pr}
P({\bf r})=a{\rm e}^{-\frac{b}{2}r^2}.
\end{equation}
which gives 
\begin{equation}
\label{norma}
{\cal N}=\frac{a}{2\pi b}.
\end{equation}
The value of the parameters $a$ and $b$ will be determined from two different
measurements from high energy proton-anti-proton scattering.

The first one, we refer to, is a measurement of the exponential slope
for the momentum transfer, $t$, in elastic $\rm \bar{p}p$ scattering.
In our model this is described by the convolution 
of the two-pomeron form-factor, $\phi_{4({\cal P})}$, with itself. With the 
approximative form in eq. (\ref{fapp}), the transverse momentum distribution
in pomeron exchange is then given by $P^2(r)\propto {\rm e}^{-br^2}$. Therefore,
our parameter $b$ is directly given by the exponential slope in 
$\rm \bar{p}p$ scattering.

This slope varies however weakly with the total energy. We will use the 
result \cite{UA4}
from UA4 at CERN, where protons (and anti-protons) were scattered at 
$\sqrt{s}=546~{\rm GeV}$.
This energy is perhaps most relevant for the HERA experiment. The
measured slope actually varies with the value of $t$, it becomes flatter for
higher values of $-t$. Since we want to fix a normalization, the low-$t$
region, which dominates, should be more relevant. In \cite{UA4}, a fit is made
to the measured distribution, and in the region
$-t<0.15~{\rm GeV}^2$, the result for the slope is \footnote{We are making
  the approximation $r^2=-t$. This is of course a very good 
estimate.}
$b=15.2\pm0.2{\rm GeV}^{-2}$.

The other measurement is that of the so called "effective" cross-section 
$\sigma_{\rm eff}$ in $\rm \bar{p}p$ scattering. This cross-section can be defined
from the cross-section $\sigma_2$ for having two independent 
high-$p_\perp$ scatterings (Resulting in four uncorrelated high-$p_\perp$ jets.)
in a single $\rm \bar{p}p$ collision. The effective cross-section is then
given by a relation between $\sigma_2$ and the square of the {\em inclusive} 
cross-section for a single high-$p_\perp$ scattering, according to
$$
\sigma_{\rm eff}=\frac{1}{2}\frac{\sigma_1^2}{\sigma_2}
$$
In practice, it turned out to be experimentally much more efficient to
measure $\sigma_{\rm eff}$ from events, with one of the four jets, replaced by a
photon. This measurement was made by CDF \cite{CDF}, and gave the result
$\sigma_{\rm eff}=14.5\pm1.7^{+1.7}_{-2.3}$~mb. 

In our model, the cross-section for one gluon exchange, is given, to leading
order, by\footnote{We have simplified the notation considerably by suppressing
  dependencies on the longitudinal variables. The final result is not affected
  by this.}
$$
\sigma_1^{(2)}=\phi_2\otimes\phi_2=
g^4\frac{N^2}{N^2-1}
\int \frac{d^2{\bf k}}{(2\pi)^2k^4}n^2(k)
$$
In this case, we can neglect the higher order contribution, 
$\sigma_1^{(4)}$ since $\sigma_1$ is the {\em inclusive} cross-section 
for one scattering.

We can now determine the cross-section for double scattering via
$\sigma_2=2\sigma_0$ and
\begin{eqnarray}
\sigma_0&=&\frac{1}{4}\phi_4\otimes\phi_4=
\frac{g^8}{4}\left(\frac{N^2}{N^2-1}\right)^2 \times \\
&\int& \left[f_4({\bf p}_1+{\bf r},-{\bf p}_1;
{\bf p}_2-{\bf r},-{\bf p}_2)\right]^2
\frac{d^2{\bf p}_1}{(2\pi)^2({\bf p}_1+{\bf r})^2p_1^2}
\frac{d^2{\bf p}_2}{(2\pi)^2({\bf p}_2-{\bf r})^2p_2^2}
\frac{d^2{\bf r}}{(2\pi)^2} \nonumber
\end{eqnarray}
where we have made the substitution in eq. (\ref{parsub}). With the assumption
that $p_1$ and $p_2$ are much larger than $r$, we can make the approximation
in eq. (\ref{fapp})
\begin{eqnarray}
\sigma_0&=&\frac{g^8}{4}\left(\frac{N^2}{N^2-1}\right)^2
\int n^2(p_1)n^2(p_2)P^2(r)
\frac{d^2{\bf p}_1}{(2\pi)^2p_1^4}
\frac{d^2{\bf p}_2}{(2\pi)^2p_2^4}
\frac{d^2{\bf r}}{(2\pi)^2} \nonumber \\
&=&\frac{1}{4}
\sigma_1^2\int P^2(r)\frac{d^2{\bf r}}{(2\pi)^2}
\end{eqnarray}

This leads to the simple relation
\begin{equation}
\sigma_{\rm eff}=\frac{1}{2}\frac{\sigma_1^2}{\sigma_2}=
\frac{1}{4}\frac{\sigma_1^2}{\sigma_0}=
\frac{1}{\int P^2(r)\frac{d^2r}{(2\pi)^2}}
\end{equation}
and it gives us the second relation, for the determination of $a$
\begin{equation}
\label{rel2}
\frac{1}{\sigma_{\rm eff}}
={\int P^2(r)\frac{d^2r}{(2\pi)^2}}=\frac{a^2}{4\pi b}
\end{equation}

From Eqs. (\ref{rel2}) and (\ref{norma}), and with the values 
$b\approx15~{\rm GeV}^{-2}$ and 
$\sigma_{\rm eff}\approx 15~{\rm mb}\approx 37~{\rm GeV}^{-2}$
we get $a\approx 2.3$, and the normalization factor
$$
{\cal N}=
\frac{1}{\sqrt{\pi b \sigma_{\rm eff}}}
\approx 0.024{\rm GeV}^2
$$

Now, we have everything we need for the calculation of the relative effect of
double gluon exchange, on the distribution of the momentum transfer in
$\gamma^*$-proton scattering.

\section{Numerical results for the momentum-transfer}
\label{num}
In this section, we will present some numerical results from the model for
double gluon exchange, that we have developed. We will consider $\gamma^*$-proton
scattering, and make an estimate of the relative effects from double gluon
exchange, on the distribution of transverse momentum, that is exchanged
between the proton and the quark pair that originates from the virtual photon.

More specifically, we will make an estimate of the function
\begin{equation}
\label{Delta}
\Delta(p_\perp^2)\equiv
\left(
\frac{d\sigma_2}{dp_\perp^2}+\frac{d\sigma_1^{(4)}}{dp_\perp^2}
\right)
\left/
\frac{d\sigma_1^{(2)}}{dp_\perp^2}\right.
\end{equation} 
We will now write the formulas for the three differential cross-sections in
detail. For the ${\cal O}(\alpha_s^2)$ contribution to one gluon exchange, we
have, from Eqs. (\ref{12diff}) and (\ref{f2n})
\begin{equation}
\frac{d\sigma_1^{(2)}}{dp_\perp^2}=
4N\pi\alpha_s^2
D_2(p_\perp)\frac{1}{p_\perp^4}n(p_\perp)
\end{equation}

For the contribution from double gluon exchange, we get from eq. (\ref{s2}) and
the substitution in eq. (\ref{parsub})
\begin{eqnarray}
\frac{d\sigma_2}{dp_\perp^2}&=&
\frac{\pi}{4}g^8N\frac{1}{(2\pi)^6}
\int \frac{d^2{\bf p}_1}{({\bf p}_1+{\bf r})^2p_1^2}
\frac{d^2{\bf p}_2}{({\bf p}_2-{\bf r})^2p_2^2}d^2{\bf r} 
\delta ({\bf p}_1+{\bf p}_2-{\bf p}_\perp)
\times\\
&\times&
f_4({\bf p}_1+{\bf r},-{\bf p}_1;{\bf p}_2-{\bf r},-{\bf p}_2)
D_{(4;0)}^{(+)}({\bf p}_1+{\bf r},
-{\bf p}_1,{\bf p}_2-{\bf r},-{\bf p}_2) \nonumber \\
&\approx&
4N\pi\alpha_s^4  \frac{\cal N}{p_\perp^4}
\int \frac{d^2{\bf p}_1}{p_1^4}\frac{p_\perp^4}{({\bf p}_\perp-{\bf p}_1)^4}
n(p_1)n({\bf p}_\perp-{\bf p}_1) \times\nonumber\\
&\times&
D_{(4;0)}^{(+)}({\bf p}_1,-{\bf p}_1,
{\bf p}_\perp-{\bf p}_1,{\bf p}_1-{\bf p}_\perp) \nonumber
\end{eqnarray}
where we have made the approximation in eq. (\ref{fapp}) and used the
definition in eq. (\ref{norm}). 
The normalization constant has the value ${\cal N}\approx 0.024$ ${\rm
GeV}^2$. 
In the same way, we can write for the ${\cal O}(\alpha_s^4)$ correction to the
single gluon process\footnote{This is by no means all of the ${\cal
    O}(\alpha_s^4)$ corrections to the one gluon exchange process (Not even,
  in the asymptotic limit). The idea is that, it is all of
  the ${\cal O}(\alpha_s^4)$ correction, in the asymptotic limit, which is 
proportional to $n^2$.} eq. (\ref{s1})
\begin{eqnarray}
\frac{d\sigma_1^{(4)}}{dp_\perp^2}&=&
-\frac{\pi}{2}g^8N\frac{1}{(2\pi)^6}
\int \frac{d^2{\bf p}_1}{({\bf p}_1+{\bf r})^2p_1^2}
\frac{d^2p_2}{({\bf p}_2-{\bf r})^2p_2^2}d^2{\bf r} 
\delta ({\bf p}_1-{\bf p}_\perp)
\times\\
&\times&
f_4({\bf p}_1+{\bf r},-{\bf p}_1;{\bf p}_2-{\bf r},-{\bf p}_2)
D_{(4;0)}^{(+)}({\bf p}_1+{\bf r},
-{\bf p}_1,{\bf p}_2-{\bf r},-{\bf p}_2) \nonumber \\
&\approx&
-8N\pi\alpha_s^4 \frac{\cal N}{p_\perp^4}
\int \frac{d^2{\bf p}_2}{p_2^4}
n(p_\perp)n(p_2)
D_{(4;0)}^{(+)}({\bf p}_\perp,-{\bf p}_\perp,{\bf p}_2,-{\bf p}_2) \nonumber
\end{eqnarray}

With the relation
$$
D_{(4;0)}^{(+)}({\bf p}_1,-{\bf p}_1,
{\bf p}_2,-{\bf p}_2)=
2D_{(2;0)}({\bf p}_1)
+2D_{(2;0)}({\bf p}_2)
-D_{(2;0)}({\bf p}_1+{\bf p}_2)-D_{(2;0)}({\bf p}_1-{\bf p}_2)
$$
derived from the Eqs. (\ref{d2nf}), (\ref{d402}) and (\ref{symmdef}), we can,
finally, write the result for the correction function, defined in
eq. (\ref{Delta}) 
\begin{eqnarray}
\Delta (p_\perp^2)&=&\alpha_s^2 {\cal N} 
\int \frac{d^2{\bf p}_1}{p_1^4}\frac{n(p_1)}{D_{(2;0)}(p_\perp)}
\left\{
\frac{p_\perp^4}{n(p_\perp)}
\frac{n({\bf p}_\perp-{\bf p}_1)}{({\bf p}_\perp-{\bf p}_1)^4}\right.
\times\\
\times
\left[
2D_{(2;0)}({\bf p}_1)
\right. &+& \left. 
2D_{(2;0)}({\bf p}_\perp-{\bf p}_1)
-D_{(2;0)}({\bf p}_\perp)-D_{(2;0)}(2{\bf p}_1-{\bf p}_\perp)
\right]
\nonumber\\
-2\left[
2D_{(2;0)}({\bf p}_1)
\right. &+& \left.\left. 
2D_{(2;0)}({\bf p}_\perp)
-D_{(2;0)}({\bf p}_\perp+{\bf p}_1)-D_{(2;0)}({\bf p}_\perp-{\bf p}_1)
\right]
\right\}.
\nonumber
\end{eqnarray}

\begin{figure}
\begin{picture}(0,245)
\Large
\put(5,225){$\Delta$}
\normalsize
\put(26,210){0.2}
\put(27,163){0.1}
\put(36,116){0}
\put(18,69){$-0.1$}
\put(17,24){$-0.2$}
\put(43,15){0}
\put(96,15){10}
\put(155,15){20}
\put(212,15){30}
\put(268,15){40}
\put(325,15){50}
\put(-5,2){\psfig{figure=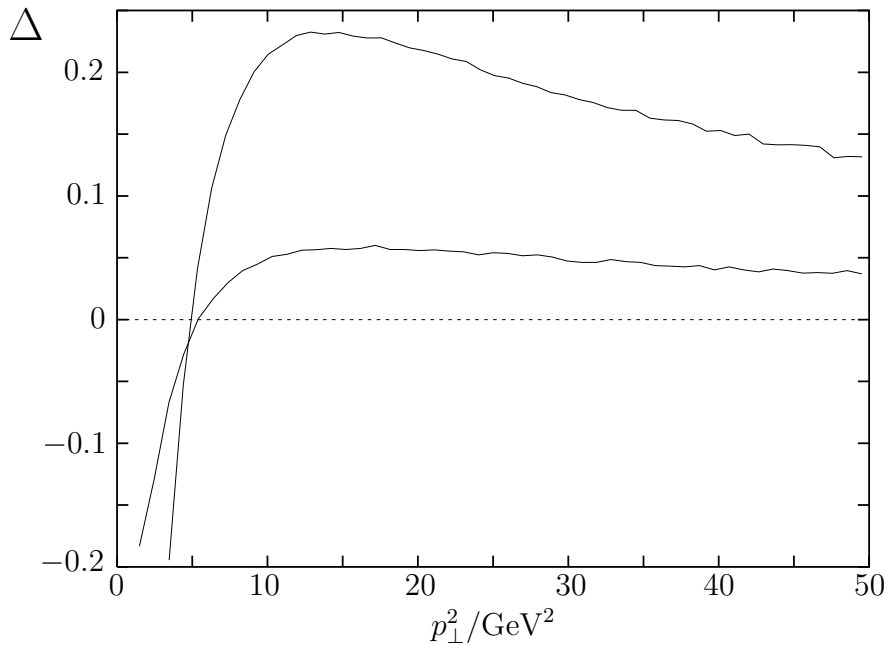,width=12cm}}
\put(165,0){$p_\perp^2/{\rm GeV}^2$}
\end{picture}
\caption{Upper and lower estimates for the correction function 
$\Delta(p_\perp^2)$, (Eq. (\ref{Delta})).}
\label{ptd}
\end{figure}

In Fig. \ref{ptd}, two different estimates of the correction function 
$\Delta(p_\perp^2)$ are plotted
for electron-proton DIS at $x_B=0.0001$ and ${\rm Q}^2=10$ ${\rm GeV}^2$. The
integration over ${\bf p}_1$ was done numerically. For the two-gluon impact factor
$D_{(2;0)}$ (See eq. (\ref{d2an})) the $y$ integration was done analytically,
while, for the $\alpha$ dependence, the constant value $\alpha(1-\alpha)=1/6$ was
inserted\footnote{The average value of $\alpha(1-\alpha)$ in the range
  $[0:1]$. This is, of course, an approximation.}. The larger estimate of
$\Delta$ in Fig. \ref{ptd}, is given by $\lambda=0.4$ and a constant
coupling $\alpha_s=0.3$. The smaller estimate is from the choices 
$\lambda=0.3$ and $\alpha_s=0.2$. 

The lower choice for the coupling, corresponds to a scale of the
order 20~${\rm GeV}^2$. Though we are interested in the distribution at
rather large $p_\perp$ values, it seems likely that the correction to the 
$p_\perp$-distribution is mainly due to an additional rather soft gluon, with
a transverse momentum of the order 1~${\rm GeV}^2$. In case, this is the 
correct scale as the argument of $\alpha_s$ for the second gluon, the 
larger value for the coupling, would be more appropriate.

We see in fig. (\ref{ptd}) that the correction function has its maximum at
around 15~${\rm GeV}^2$ with the maximal corrections of 6~$\%$ and 23~$\%$,
respectively, in the two estimates. Towards higher
$p_\perp$ values, the correction decreases, but continues to be positive. For
the low $p_\perp$ values, there is a large negative correction. In this
region, the correction from three-gluon exchange should be positive. In the
case that the negative corrections in this region are large (At the order of
one, or greater) they could not be trusted since the positive correction from
three gluon exchange would also be large. When it comes to the negative
corrections in the low-$p_\perp$ region, one is, for any treatment with a
fixed number of exchanged gluons, always dependent on some kind
of lower cut-off.

These negative corrections, at the order of one, are however, a signal for an
interesting phenomenon that have been expected from other models (See e.g. 
\cite{MC}), namely, saturation in the transverse momentum distribution. If
multiple gluon exchange to all orders, are taken into account, one expects
that the, $p_\perp$-distribution, which is peaked at zero for one gluon
exchange, will instead become flat below a lower limit, $\Lambda$. One can
make the interpretation, that the density of gluons is so large, in this 
region, so that in almost every scattering event, at least one of them is
absorbed by the quarks. The limit, $\Lambda$, is expected to grow, towards
lower $x_B$-values \cite{MC}.

We will now list the main approximations  that have been used and that can
affect our result for the $p_\perp$ distribution. The severity of each of
these approximations is estimated by the factor, $\rho$, that the (positive) 
maximal value of the correction function, $\Delta$ is changed, if we would do the 
approximation, in another, but still reasonable way. In this respect, we will
refer the lower estimate as the "Default" choice (See Table \ref{etab}.).

There are three entities that are directly proportional to the final result.
These are the normalisation factor $\cal N$, the square of the coupling
constant $\alpha_s^2$ and the number of resolved partons $n(k)$.

For the normalisation factor, $\cal N$, the experimental uncertainty in
$\sigma_{\rm eff}$ of about 10~\%, will result in the same relative error in
$\Delta$. An error, perhaps equally important, should arise from our
simple physical assumptions in the determination of $\cal N$. Further, 
the two choices of $\alpha_s=0.2$ and $\alpha_s=0.3$ give a relative factor 
$\rho=2.25$.

The discussion of errors on the number of partons $n(k)$ is more involved,
but important. From Eqs. (\ref{hatn}) and (\ref{nk}), one can see that there
are three parameters that determine $n(k)$. These are the exponent 
$\lambda$, the gluon density at $x_p=0.1$ (This was set to 2) and the 
scale at which, coherence effects in the proton sets in, 
$\mu^2=0.1$~${\rm GeV}^2$.

The scale $\mu^2$ has rather small effects on the maximal (positive) value 
of $\Delta$. An increase to the value $\mu^2=0.5$~${\rm GeV}^2$, reduced the
maximal positive correction from $\sim 6\%$ to $\sim 4\%$ ($\rho =0.67$). 
The maximum was shifted to larger $p_\perp$ values, with 
approximately 5~${\rm GeV}^2$.

For the dependence on $\lambda$, by changing the value from 0.3 to 0.4, 
the number of resolved gluons, in the low-$k^2$ region is more than doubled
(See Fig. \ref{pif}). This results in an doubling of the maximal
correction, that is $\rho=2$.

There is another uncertainty on our gluon density $n(k)$, which is
related to the fact that we have used a formalism which is valid in the asymptotic 
$s\rightarrow\infty$ limit. In the derivation of $n(k)$, we introduced the
lower limit $\hat{x}(k^2)$, for the momentum fraction,
that a resolved gluon could have. This limit was derived from a local
invariant energy squared $\hat{s}$ and we required, for the exchanged gluon
that $k^2<\hat{s}/4$. One should expect that, compared to the asymptotic
result, there is a suppression also, in a range below this value (In the case
of time-like cascades, see e.g. \cite{lund}). 
To estimate the significance of
this approximation, the function $n(k)$ was reevaluated with the limit
$k^2<\hat{s}/8$. This lead to a reduction of $n(k)$ of about 20~$\%$, over
the whole range, except for $p_\perp$ values below 2~${\rm GeV}^2$, where $n$
was not changed. In this region, the limit $\hat{x}$ is insignificant because
it becomes smaller than $x_B$. The effect on the correction function $\Delta$, 
was that it had the same shape and maximal value, but was shifted to lower
$p_\perp$ values, a distance of 5~${\rm GeV}^2$. So, in this case,
we had $\rho =1$.

It might be interesting to check how the maximal value of the correction,
$\Delta$, varies with the values of $x_B$ and $Q^2$. Therefore, separate runs
were made for ($x_B$, $Q^2$) = ($10^{-3}$, 10~${\rm GeV}^2$), 
($10^{-5}$, 10~${\rm GeV}^2$),  ($10^{-4}$, 20~${\rm GeV}^2$) and 
($10^{-4}$, 5~${\rm GeV}^2$) (See Table \ref{etab} for the results for
$\rho$). The conclusion is that the effect of double gluon exchange increases
rapidly towards, lower values of both $x_B$ and $Q^2$. Another observation
from these runs is that the maximum always occured at $p_\perp^2$ values
slightly above $Q^2$. The $Q^2$-dependence of the maximal correction, seems to
be consistent with the expectation that multiple interactions is a
higher twist effect. A $1/Q^2$-behaviour could also be explained by the fact
that $\Delta$ is dimension less, but proportional to $\cal N$ which has the
dimension $[{\rm energy}]^2$. (The dependence on $\mu^2$ was observed above to
be weak, perhaps logarithmic.). 

\begin{table}[htb]
\begin{center}
\begin{tabular}{|c|c|c|}
\hline
Entity & Alternative procedure (Default)& $\rho$ \\ \hline\hline
${\cal N}$          &  
[Experimental errors] &
$\pm10~\%$
                                  \\ \hline  
${\cal N}$          &  
[Theory assumptions] &
$\pm10~\%$
                                  \\ \hline  
$\alpha_s$     &  
Set to 0.3 (0.2)    &
$2.25$  
                                  \\ \hline  
$n$          &  
$\mu^2=0.5$~${\rm GeV}^2$ (0.1~${\rm GeV}^2$)      &
0.67
                                  \\ \hline  
$n$          &  
$\lambda=0.4$ ($\lambda=0.3$) &
$2$  \\  \hline

$n$          & 
$\hat{x}$ determined by $k^2\!<\!\hat{s}/8$ ($k^2\!<\!\hat{s}/4$)   &
$1$   \\ \hline \hline 

$x_B$          & 
$10^{-5}$ ($10^{-4}$) &
$1.9$   \\ \hline

$x_B$          & 
$10^{-3}$ ($10^{-4}$) &
$0.5$   \\ \hline

$Q^2$          & 
5~${\rm GeV}^2$ (10~${\rm GeV}^2$) &
$1.8$   \\ \hline

$Q^2$          & 
20~${\rm GeV}^2$ (10~${\rm GeV}^2$) &
$0.5$   \\ \hline

\end{tabular}
\caption{Estimate of various uncertainties in the final answer. $\rho$ is the
  relative effect on the maximal correction. In the second part of the table,
  the dependence on $x_B$ and $Q^2$, is displayed.}
\label{etab}
\end{center}
\end{table}

There have been several measurements and analyses of jet-distributions at
HERA, in the region of low $x_B$ and $Q^2>1$~${\rm GeV}^2$, that is relevant 
for the discussion here (E.g. \cite{Jets}). The high-$p_\perp$ jet activity, 
has been described
by the definition of jet-rates: The fraction of the cross-section, with  a
number of jets above some high $p_\perp$-value. There has also been specific
studies of the fraction of the photon energy, that is carried in these jets.
In this way, one determines if it is more likely that the jet originates
from a quark (high energy fraction) or a gluon, which is more likely to be
in the proton direction (low fraction of the photon energy).

The result, for the momentum transfer, that has been presented here, can not
be directly compared with the measured jet-distributions. The main reason is that
the transverse momenta of the quarks was integrated out. It is, of course, not
necessary to do this, but the $p_\perp$-distribution 
of the individual quarks, in the high-$p_\perp$ region, is very much dependent
on the energy that is available for the quark-anti-quark system. In the
leading log(s) formalism, that we have used, there is no information on this
energy (It is assumed to be infinite), and one would have to determine it, by
other arguments\footnote{This means that we have overestimated the 
  cross-sections, in the treatment above, but this error is small 
  inclusively over the quark momenta and should rather well factor out 
  in the relative correction $\Delta$.}.

So far, there has not been any direct measurements of the momentum transfer to
the quark pair. In the case of very high momentum transfers, 
$p_\perp^2>Q^2$, one could, however, be rather certain, that this will lead to
two jets, with transverse momenta $\sim p_\perp$, originating from one of the
quarks, and the gluon which takes the recoil (The largest recoil, in the case
of double gluon exchange). The other quark, with a lower transverse momentum, 
should then be observable as the photon remnant. Our correction to the momentum
transfer, would therefore lead to similar corrections (enhancements) in the so
called "resolved-photon" events, with high $p_\perp$.

Another interesting observation, that one could perhaps make at HERA, is 
related to the negative correction, in the region of low momentum transfers. 
If the saturation limit ($\Lambda$ above) is sufficiently large, it should be
observable in the $p_\perp$-distribution of the quark jets, in a frame were
the proton and the photon have zero transverse momentum. This would be 
possible, only if
$\Lambda$ is large enough, in the region of the lowest $x_B$-values, 
available at HERA, so that the effect is not drowned by the hadronisation 
process. From the larger estimate, above, this saturation should be expected
to occur at $\Lambda^2 \sim 2$~${\rm GeV}^2$. 

\section{Summary}
\label{S}
We have studied the process of double gluon exchange, in non-diffractive 
$\gamma^*$-proton DIS. A new model was developed, based on previous 
descriptions of diffractive events, as the perturbative
exchange of two gluons, originating from one pomeron \cite{Diff}. The 
relation to the non-diffractive process, was determined from a general 
result from Regge theory: The AGK cutting rules \cite{AGK}. The validity 
of this procedure has,
however, not been fully proven (In the leading log(s) expansion, these
contributions are next-to-next-to-leading.). 

As a first application of the model, we studied the corrections to the
transverse momentum transfer to the quark-anti-quark pair, at the 
photon vertex. We found significant enhancements (At the order of 10\%) of the
cross-section for large momentum transfers, and large negative
corrections, ($\sim 1$) for momentum transfers close to zero. The prediction
for the negative corrections is however not reliable, but signals saturation
in the transverse momentum distribution, in an all orders (in the number 
of exchanged gluons), treatment. The possibility of the observation of these
effects in the measured jet-distributions at HERA, was discussed.

The results presented here, should be taken as order of magnitude estimates,
and a description of the general behaviour. It should be possible to make the
model more accurate, in order to make direct comparisons with data. Some of
the necessary improvements, like a new fit of the parton distributions, are
straight forward, but involved. The most effective way of doing this
would probably be to include the perturbative multiple interaction as a new 
feature in existing models for DIS.

\section*{Acknowledgements}
Many people have been kindly helpful with comments and suggestions. 
Among them are Joachim Bartels, Torbj\"orn Sj\"ostrand and Anders Edin.    

\appendix
\section*{A consistency check}
In section \ref{logs}, we could make use of the Cutkoski rule since the
amplitude for pomeron exchange is purely imaginary to leading order. This 
is expected for amplitudes which describe the exchange
of a symmetric colour state.
We will now, with the same method as $\sigma_0$ was calculated, calculate the
contribution to $\sigma_2$ which is due to just the imaginary part of the
amplitude for two gluon exchange. For the exchange of symmetric colour states,
this is the leading contribution. Consequently, our method should give the
correct AGK relation $\sigma_2:\sigma_0=2:1$, for the parts of $\sigma_2$
which are due to the exchange the symmetric colour states and the
corresponding part of $\sigma_0$. The main argument is that both $\sigma_0$
and the symmetric part of $\sigma_2$, have their contributions given from the
symmetric part, $D^{(+)}({\bf k}_1,{\bf k}_2,{\bf k}_3,{\bf k}_4)$ (Defined in
Eqs. (\ref{symmdef}) and (\ref{d402})), of the photon impact factor, and that 
this function is invariant under permutations of its arguments. 

To do this, we need the full impact factor of the proton (Fig. \ref{proto}), 
whereas, before, we used only the part with one pomeron-proton vertex on 
each side of the final-state cut. We choose a form which is explicitely 
symmetric in the simultaneous exchange of the colour and momentum of any 
two gluons.
\begin{eqnarray}
\label{phi4}
\phi_4^{a_1a_2a_3a_4}({\bf k}_1,{\bf k}_2,{\bf k}_3,{\bf k}_4)&=& \\
\phi_{4({\cal P})}^{a_1a_2a_3a_4}({\bf k}_1,{\bf k}_2,{\bf k}_3,{\bf k}_4)&+&
\phi_{4({\cal P})}^{a_1a_3a_2a_4}({\bf k}_1,{\bf k}_3,{\bf k}_2,{\bf k}_4)+
\phi_{4({\cal P})}^{a_1a_4a_3a_2}({\bf k}_1,{\bf k}_4,{\bf k}_3,{\bf k}_2)
\nonumber
\end{eqnarray}
with $\phi_{4({\cal P})}$ defined in eq. (\ref{p0}).

When the impact factor above is convoluted with the photon's impact factor, 
we will get three terms, arising from the three terms in $\phi_4$. The first of
these is what was discussed before, the cross-section $\sigma_0$, while the
sum of the other terms is the contribution to $\sigma_2$. As discussed before,
for the contributions to $\sigma_2$ which are due to the exchange of symmetric
colour states, we can again make use of the multiply cut photon impact factor,
but now, convolute it with the second and third terms of $\phi_4$. 
To do this, we have to make projections into the symmetric colour states. In 
the case of $\sigma_2$, we
can do this by putting in a symmetry projector on the colours of gluon one and
two. This part of the impact factor has already been singled out in
eq. (\ref{separation}). We can now see that, in the convolution with the
second or third terms of $\phi_4$, we get the same colour factor:
$$
d_+^{abcd}\delta_{ac}\delta_{bd}=
d_+^{abcd}\delta_{ad}\delta_{bc}=
\left( C_f-\frac{1}{2N}\right)NC_f
$$
Further, the two terms have the factors 
$$D^{(+)}({\bf k}_1,{\bf k}_2,{\bf k}_3,{\bf k}_4)
f({\bf k}_1,{\bf k}_3;{\bf k}_2,{\bf k}_4)$$ and 
$$D^{(+)}({\bf k}_1,{\bf k}_2,{\bf k}_3,{\bf k}_4)
f({\bf k}_1,{\bf k}_4;{\bf k}_3,{\bf k}_2)$$ respectively. Since $D^{(+)}$ is
invariant under permutations of the momenta, these two factors are equal to
each other (After renumbering) and also equal to the corresponding factor 
which appears for $\sigma_0$. 

The corresponding colour factor for $\sigma_0$ is
$$
d_+^{abcd}\delta_{ab}\delta_{cd}=
2NC_f^2
$$
From the AGK rule, this would mean that the total contribution to $\sigma_2$
should have the factor $4NC_f^2$. Above, we have calculated the
contribution from the exchange of symmetric colour states. Consequently, 
the validity of the
AGK rule, gives for the anti-symmetric exchange part of $\sigma_2$, that it
should come with the same momentum dependence, and with the colour factor
$$
4NC_f^2-2\left( C_f-\frac{1}{2N}\right)NC_f=N^2C_f
$$

We will now calculate the colour factor for the part of $\sigma_0$
which corresponds to the symmetric colour exchange in $\sigma_2$~\footnote{It
 is acknowledged that the exact definition of this procedure is 
perhaps not known.}.
For $\sigma_0$, the colour symmetry projector should be connected with  one 
gluon from each pomeron, e.g. the gluons 2 and 3 or 1 and 3. There is also a second
requirement, namely that the two colour indices that are symmetrised over are
directly connected to each other through one quark line. This is the case for 
the symmetrisation procedure that was used for the $\sigma_2$ contribution:
$$
d^{a_1a_2a_3a_4}\rightarrow 
d_+^{a_1a_2a_3a_4}\equiv\frac{1}{2}\left(d^{a_1a_2a_3a_4}+d^{a_2a_1a_3a_4}\right)
$$
The symmetrisation is done over neighbouring cites of the tensor $d$.

For $\sigma_0$, this can be done in the following way for the first term in
Eq. (\ref{d401})
$$
d^{a_1a_2a_3a_4}\rightarrow 
\frac{1}{2}\left(d^{a_1a_2a_3a_4}+d^{a_1a_3a_2a_4}\right)
$$
and for the second term
$$
d^{a_2a_1a_3a_4}\rightarrow 
\frac{1}{2}\left(d^{a_2a_1a_3a_4}+d^{a_2a_3a_1a_4}\right)
$$
After this procedure, and multiplying with $\delta_{a_1a_2}\delta_{a_3a_4}$,
we get the same colour factor, $(C_f-1/2N)NC_f$, also for the $\sigma_0$ 
contribution.

Thus, after the symmetrisations, we get the same contribution to $\sigma_0$ 
from the first term of $\phi_4$, as each of the contributions from the other
two terms to $\sigma_2$. That is, the symmetrised contributions satisfy the
AGK relation $\sigma_0:\sigma_2=1:2$. This result depends on the form of
$\phi_4$ that has been used, but we have shown that it is possible to
choose a form, inspired by an underlying picture of two pomerons that couple
in all possible ways, so that the AGK rule is satisfied.

\end{document}